%% file: ms.tex
\documentclass[
  aps,
  prb,
  reprint,
  superscriptaddress,
  longbibliography,
  showkeys,
  citeautoscript,
]{revtex4-1}
\input{settings}

\newcommand{\dtu}{
    Department of Electrical and Photonics Engineering, Technical University of Denmark,
    2800 Kgs. Lyngby, Denmark.
}

\begin{document}

\title{Tunable and low-noise WSe$_2$ quantum emitters for quantum photonics}

\author{Athanasios Paralikis}
%\altaffiliation{Contributed equally to this work}
\affiliation{\dtu}

\author{Paweł Wyborski}
%\altaffiliation{Contributed equally to this work}
\affiliation{\dtu}

\author{Pietro Metuh}
\affiliation{\dtu}

\author{Niels Gregersen}
\affiliation{\dtu}

\author{Battulga Munkhbat}
\email[]{bamunk@dtu.dk}
\affiliation{\dtu}

\begin{abstract}

Low-noise and tunable single-photon sources are essential components of photonic quantum technologies. 
However, in WSe\textsubscript{2} quantum emitters, charge noise from fluctuations in their local electrostatic environment remains a major obstacle to achieving transform-limited single-photon emission and high photon indistinguishability.
Here, we systematically investigate two noise mitigation strategies in hexagonal boron nitride (hBN) encapsulation and electrostatic biasing. We demonstrate that hBN encapsulation alone suppresses spectral wandering (from $\sim$170~$\mu$eV  to $\sim$40~$\mu$eV) and narrows emission linewidths (from $\sim$500~$\mu$eV to $\sim$150~$\mu$eV), while applied bias enables stable Stark tuning over a 280~$\mu$eV range and further linewidth narrowing down to $\sim$100~$\mu$eV reaching the resolution-limited regime. Time-resolved and second-order correlation measurements confirm stable mono-exponential decay and high single-photon purity ($g^{(2)}(0) \approx 0.01$) with no observable blinking.
To quantify progress toward the transform limit, we define two figures of merit: the linewidth ratio $R = W_{\text{exp}} / W_{\text{rad}}$ and total broadening $\Delta W = W_{\text{exp}} - W_{\text{rad}}$, with both being reduced more than five-fold in optimized devices. 
These results provide a robust framework for developing and evaluating low-noise, tunable WSe\textsubscript{2} quantum emitters, potentially realizing electrically controllable sources of indistinguishable single-photons for future photonic quantum technologies.

\end{abstract}


\maketitle

\section{Introduction}
High-performance single-photon sources (SPSs) are fundamental building blocks for photonic quantum technologies,\cite{gregersen_mørk_2017, gao2012observation, wang2020integrated, xu2020secure} enabling a wide range of applications in quantum communications and quantum information processing. For most applications -- particularly those involving quantum interference such as photonic logic or entanglement swapping -- photon indistinguishability is a critical requirement. III–V semiconductor quantum dots have demonstrated near-unity purity and indistinguishability when integrated with optical cavities and operated under advanced excitation schemes.\cite{senellart2017high, ding2016demand, thomas2021bright, karli2024controlling, boos2024coherent, wei2022tailoring} However, their reliance on advanced epitaxial growth techniques and complex nanofabrication poses challenges for large-scale deployment and integration.

In recent years, two-dimensional (2D) semiconductors, particularly transition metal dichalcogenides (TMDs), have emerged as a highly promising platform for developing on-demand single-photon sources.\cite{montblanch2023layered, So+2025, esmann2024solid} These materials offer several unique advantages, including large exciton binding energies, valley degrees of freedom, robust trion formation, and the ability to form heterostructures and Moiré excitons.\cite{liu2015molecular, montblanch2023layered, tran2019evidence, singh2016trion} Additionally, their van der Waals (vdW) stacking nature allow for their direct integration to a wide range of photonic nanostructures and cavities.\cite{tonndorf2017chip}

Single-photon emission in TMDs typically originates from localized excitonic states,\cite{Aharonovich2016Solid-stateEmitters, linhart} which can be developed through a variety of approaches. These include imaginative strain-engineering techniques (AFM indentation\cite{stevens2022enhancing, abramov2023photoluminescence}, nanostructures\cite{paralikis2024tailoring, Parto2021DefectK,peng2020creation, azzam2023purcell, luo2018deterministic}, nanoparticles\cite{xu2024sub, xu2023conversion, tripathi2018spontaneous}, piezoelectric substrates\cite{iff2019strain}) and deterministic defect creation via high-energy irradiation (e.g., electron\cite{paralikis2024tailoring, Parto2021DefectK, xu2023conversion}, helium ion\cite{klein2019site, micevic2022demand}, or UV beams\cite{wang2022utilizing}).

These approaches have enabled the site-specific generation of linearly polarized sources with high single-photon purity across different TMD materials.\cite{zhao2021site, micevic2022demand, klein2019site, yu2021site, paralikis2024tailoring} Moreover, the emission energies of these quantum emitters are readily tunable via mechanical strain, electrical gating, or magnetic fields.\cite{lenferink2022tunable, chakraborty2019electrical, howarth2024electroluminescent}

Despite promising advances, reported indistinguishability values for TMD quantum emitters remain low, around 2\%.\cite{drawer2023monolayer} A key limiting factor is emission linewidth broadening, driven by mechanisms such as phonon-induced dephasing and spectral diffusion from local charge fluctuations,\cite{senellart2017high} with charge noise identified as the dominant contributor.\cite{vannucci2024single} Broadening refers to the deviation of the experimental linewidth ($W_\text{exp}$) from the ideal transform-limited value ($W_\text{rad}$), which is fundamentally set by the radiative lifetime. In TMDs, lifetimes span from sub-nanosecond to tens of nanoseconds,\cite{yu2021site, klein2019site, srivastava2015optically, kumar2015strain, paralikis2024tailoring, piccinini2024high, tonndorf2015single} due to differences in their microscopic origins and complex recombination dynamics, corresponding to $W_\text{rad}$ values of 0.01–10 $\mu$eV. In contrast, measured linewidths typically lie between several hundred and a few thousand $\mu$eV,\cite{srivastava2015optically, Parto2021DefectK, piccinini2024high, branny2017deterministic} revealing a mismatch of several orders of magnitude.

Recent studies suggest that strategies such as hexagonal boron nitride (hBN) encapsulation and electrostatic biasing can suppress these fluctuations and reduce $W_\text{exp}$ by stabilizing the local electrostatic environment.\cite{chakraborty2019electrical, lenferink2022tunable, howarth2024electroluminescent, kim2019position, klein2019site} However, a comprehensive and quantitative evaluation of how different noise mitigation approaches bring TMD quantum emitter closer to the transform limit remains an open challenge.

\begin{figure*}[hbt!]
\includegraphics[width=0.8\textwidth]{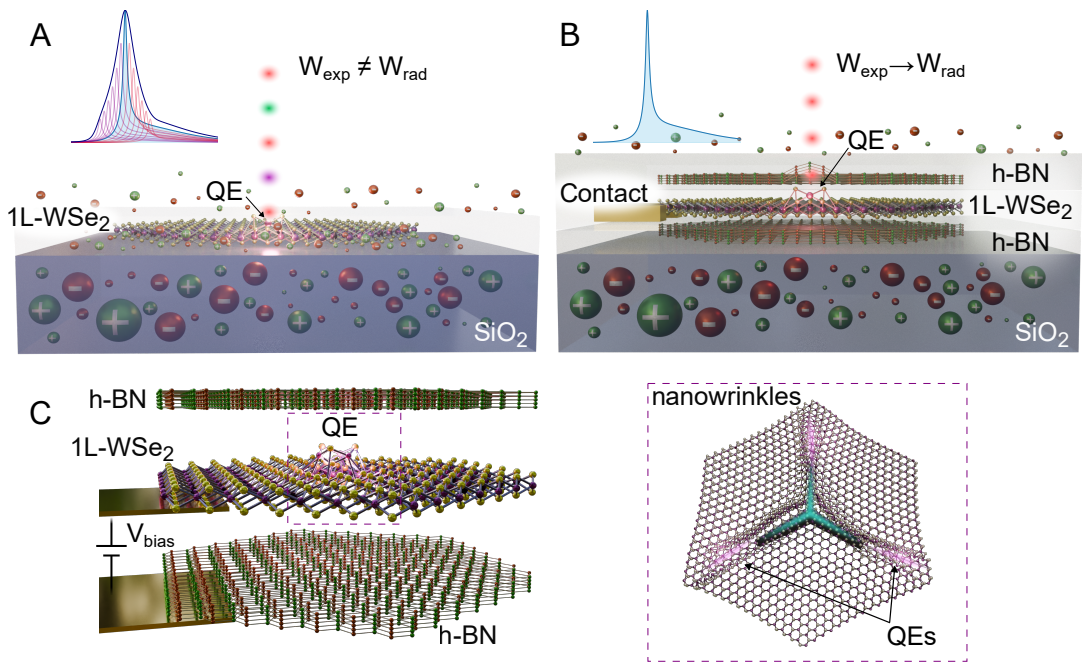}
\caption{\textbf{Schematic representation of the system under study.} (\textbf{A}) WSe\textsubscript{2} monolayer on a SiO\textsubscript{2} substrate in an unstable charge environment. The presence of fluctuating charges perturbs the quantum emitter, leading to inhomogeneous broadening through spectral diffusion, which leads to a deviation from lifetime-limited emission. This effect is illustrated by the variation in photon colors. The corresponding photoluminescence (PL) spectrum exhibits a broadened peak, where the measured linewidth (dark blue) exceeds the radiative lifetime-limited linewidth (light blue). (\textbf{B}) WSe\textsubscript{2} monolayer encapsulated in few-layer hexagonal boron nitride (hBN) on a SiO\textsubscript{2} substrate. The hBN encapsulation isolates the monolayer from the surrounding fluctuating charges, while a gold contact enables the application of an external bias to further stabilize the charge environment. As a result, the emitter experiences minimal spectral diffusion (uniform photon colors). The corresponding PL spectrum is significantly narrower, with $W_\text{exp}$ approaching $W_\text{rad}$ (lifetime-limited emission linewidth), indicating a reduction in spectral broadening effects.(\textbf{C}) Schematic representation of the fully encapsulated Device, with the inset illustrating the strain-engineered WSe\textsubscript{2} monolayer hosting the quantum emitters.}
\label{Figure1}
\end{figure*}

In this work, we systematically investigate how hBN encapsulation and electrostatic biasing can mitigate charge noise in WSe\textsubscript{2} quantum emitters. By combining passive dielectric screening with active tuning of the local electrostatic environment, we achieve significant improvements in emitter performance. We show that hBN encapsulation alone minimizes spectral diffusion, from $\sim$170~$\mu$eV down to $\sim$40~$\mu$eV, while also narrowing the emission linewidth from $\sim$500~$\mu$eV to $\sim$150~$\mu$eV. Time-resolved measurements reveal a mono-exponential decay, pointing to a more stable surrounding charge environment and less complex decay dynamics. Moreover, second-order correlation measurements confirm high single-photon purity ($g^{(2)}(0)~\sim~0.01$) with no observable bunching, indicating the absence of blinking. Electrostatic biasing further enhances performance in encapsulated devices, enabling resolution-limited linewidths in the range of $\sim$100~$\mu$eV and stable Stark tuning over a range of 280~$\mu$eV. To quantify these improvements, we introduce two figures of merit: the linewidth ratio $R = W_\text{exp} / W_\text{rad}$ and the excess broadening $\Delta W = W_\text{exp} - W_\text{rad}$. Combined encapsulation and biasing reduce both metrics more than five-fold, with $R$ decreasing from 4314~to~860 and $\Delta W$ from 474.5~to~92.8~$\mu$eV. While $\Delta W$ reaches values comparable to the state-of-the-art due to a narrow $W_\text{exp}$, $R$ remains modest, limited by the relatively long radiative lifetime ($\sim$6 ns) under above-band excitation. Nevertheless, we discuss clear routes for reaching transform-limited performance via advanced excitation schemes and Purcell-enhanced emission.
These findings establish a robust framework for developing low-noise, tunable WSe\textsubscript{2} quantum emitters, potentially enabling electrically controlled sources of indistinguishable single photons for future photonic quantum information technologies.

% ------------------------------------------------------- %

\section{Results}

\subsection{System under study}

Figure \ref{Figure1}A depicts a bare WSe\textsubscript{2} monolayer hosting a quantum emitter, on a standard SiO\textsubscript{2}/Si substrate. In this configuration, the emitter is highly susceptible to charge noise arising from trapped charges in the substrate and free carriers within the host monolayer WSe\textsubscript{2} itself. These fluctuating charges generate local electric fields that shift the exciton energy via the quantum-confined Stark effect, leading to temporal spectral fluctuations and linewidth broadening. \cite{Parto2021DefectK,daveau2020spectral,paralikis2024tailoring, piccinini2024high} As a result, the emission deviates significantly from the transform-limited regime, simultaneously reducing coherence and indistinguishability.\cite{daveau2020spectral, senellart2017high} To mitigate this charge noise, we implement two complementary strategies, as illustrated in Figure \ref{Figure1}B: \textit{i}) passive electrostatic screening through hexagonal boron nitride (hBN) encapsulation and \textit{ii}) active electrostatic control via applied electrical bias. 

To evaluate the impact of each approach, we fabricated two devices, Device A and Device B, each with their own monolayer WSe\textsubscript{2} quantum emitters formed via deterministic strain engineering on nanopillar arrays (Fig.~\ref{Figure1}C, purple inset) on typical SiO$_2$/Si substrates, as detailed in our previous work.\cite{paralikis2024tailoring} 
Device A consists of a bare WSe\textsubscript{2} monolayer directly interfaced with pre-patterned gold electrodes, enabling active biasing without dielectric isolation. In contrast, Device B features a fully hBN-encapsulated WSe\textsubscript{2} quantum emitters with similar electrical contacts, thus combining passive screening with active charge control.
It is worth noting that hBN encapsulation over strain-engineered monolayers presents specific fabrication challenges, as its high bending rigidity\cite{androulidakis2018tailoring} impedes conformal wrapping over nanopillars for nanowrinkle formation.\cite{daveau2020spectral} We addressed this by adapting the nanopillar geometry (increased height, longer tips) to accommodate a thin (2-3 layer) bottom hBN flake, followed by transfer of the WSe\textsubscript{2} monolayer and a thicker (5-8 layer) top hBN cap. Full fabrication procedures are detailed in the Materials and Methods section.

% ------------------------------------------------------- %

\begin{figure*}[t!]
\includegraphics[width=\textwidth]{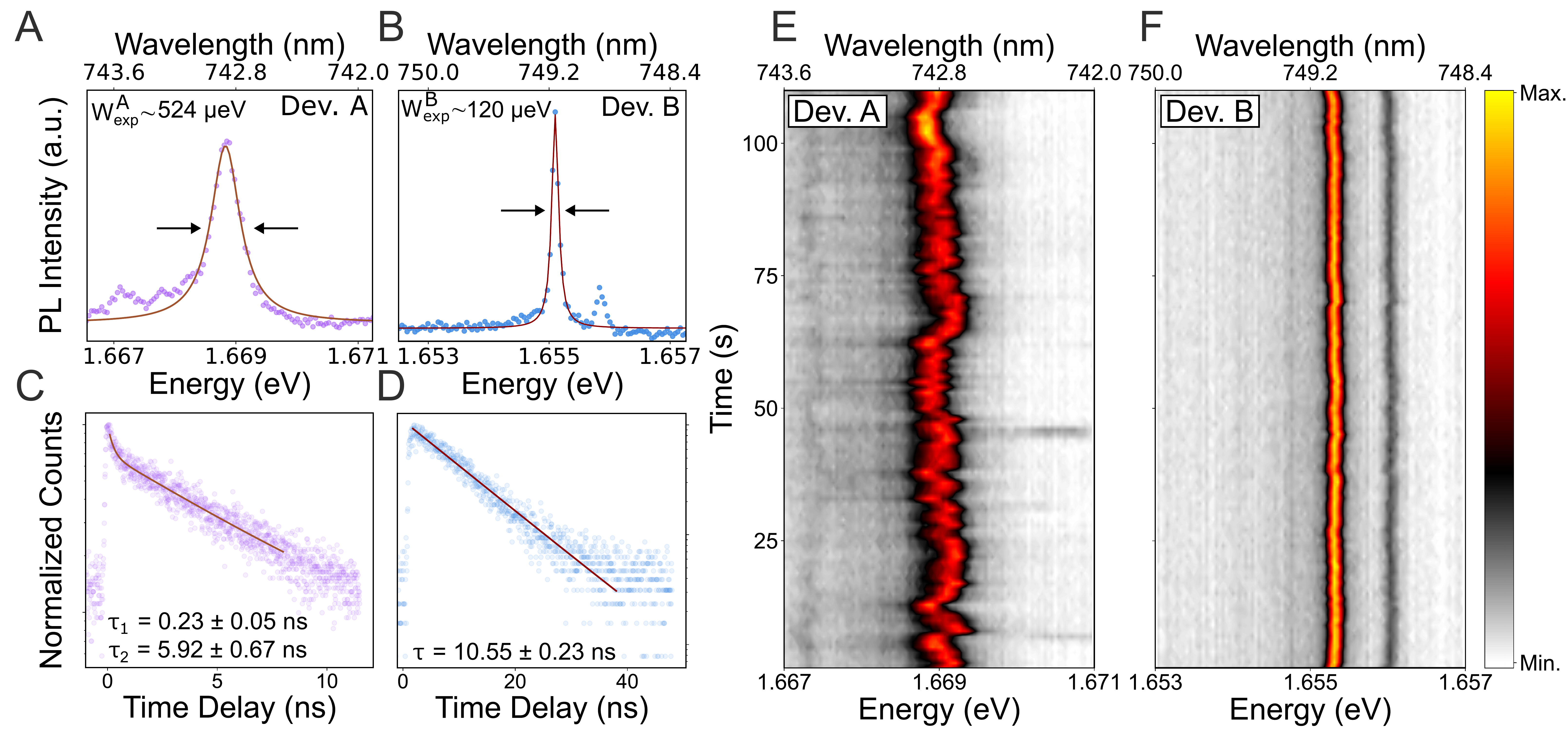}
\caption{\textbf{Comparison between unencapsulated (Device A) and encapsulated (Device B) WSe\textsubscript{2} QEs.} (\textbf{A,B}) High-resolution PL spectra at 4~K for exemplary unencapsulated (A) and encapsulated (B) quantum emitters, referred to as Devices A and B, respectively. Both spectra are fitted with a Lorentzian curve, revealing a linewidth of $0.233\pm0.008$~nm for Device A and $0.054\pm0.002$~nm for Device B, respectively. (\textbf{C,D}) Time-resolved PL measurements for Devices A and B. Device A (C) exhibits a bi-exponential decay, with an initial fast decay at $\tau_1 = 0.23\pm0.06$~ns, dominated by a long decay component of $\tau_2 = 5.92\pm0.67$~ns. Device B (D) shows simpler single-exponential decay dynamics with a calculated decay time at $\tau = 10.55\pm0.23$~ns. (\textbf{E,F}) High-resolution PL emission time trace signal for Devices A and B, collected over 110~s with an integration time of 1~s/frame. Device A (E) shows relatively unstable emission with a spectral wandering of 290~$\mu$eV around a central wavelength of $742.831\pm0.039$~nm. Device B (F) exhibits greater stability, with a spectral wandering of 70~$\mu$eV and a central wavelength of $749.072\pm0.006$~nm. All spectra are collected using a 650~nm pulsed laser excitation scheme. For the lifetime measurements, a repetition rate of 80 and 20~MHz is used for Devices A and B, respectively.}
\label{Figure2}
\end{figure*}

\subsection{Passive noise mitigation via hBN encapsulation}

First, to evaluate the impact of hBN encapsulation in mitigating charge noise, we compare representative quantum emitters from Device A (unencapsulated) and Device B (fully encapsulated) under zero applied bias. Figures~\ref{Figure2}A and B present $\mu$PL spectra recorded at T~=~4~K under 650~nm CW excitation with a $\sim$0.8~s integration time. The emitter in Device A exhibits a relatively broad emission linewidth, likely caused by spectral diffusion due to an unstable charge environment. In contrast, the emitter in Device B shows a significantly narrower linewidth, indicating reduced charge noise due to electrostatic screening provided by the hBN encapsulation. To quantify this difference, we fit the $\mu$PL spectra with Lorentzian functions and extract the experimental full width at half maximum  (FWHM), denoted as $W_\text{exp}$. Device A yields $W_\text{exp}$~=~524~$\pm$~18~$\mu$eV (0.233~$\pm$~0.008~nm) centered around 742.868~$\pm$~0.003~nm, whereas Device B shows a substantially narrower $W_\text{exp}$~=~125~$\pm$~4~$\mu$eV (0.054~$\pm$~0.002~nm) at 749.159~$\pm$~0.001~nm. This fourfold reduction in linewidth could suggest significant suppression of charge-induced broadening due to the dielectric screening provided by hBN encapsulation.

To further investigate the underlying recombination dynamics of the emitters in both Devices, we performed time-resolved photoluminescence (TRPL) measurements under 650~nm above-band pulsed excitation. Figure~\ref{Figure2}C shows the TRPL trace for Device A, which displays complex decay behavior characterized by both fast and slow components. A biexponential fit yields a dominant slow component of $\tau_2~=~5.92~\pm~0.67$~ns, in line with typical reported values for WSe\textsubscript{2}-based quantum emitters. \cite{branny2017deterministic,kumar2015strain,so2021polarization,iff2021purcell} The minor fast component of $\tau_1~=~0.23~\pm~0.06$~ns could be associated with non-radiative recombination pathways mediated by trap states and fluctuating charges in the vicinity of the emitter due to direct contact with a substrate.\cite{paralikis2024tailoring, piccinini2024high} On the other hand, Device B (Fig.~\ref{Figure2}D) exhibits a simpler mono-exponential decay with a lifetime of $\tau~=~10.55~\pm~0.23$~ns, suggesting a suppression of non-radiative processes and an enhanced stability of the local charge environment, likely due to the hBN encapsulation.

To assess the effects of hBN encapsulation on the emitters' spectral stability over longer timescales, we recorded PL time traces under 650~nm CW excitation with an integration time of $\sim$0.8~s per frame. Figures~\ref{Figure2}E and F show results for Devices A and B, respectively. Device A exhibits pronounced spectral wandering, while Device B maintains a much more stable and narrow emission, consistent with earlier observations.\cite{daveau2020spectral} To quantify this behavior, we fit each time frame with a Lorentzian function to extract the central wavelength and linewidth of the emission over time.
As shown in Supplementary Fig. S1, the fitted results reveal a pronounced contrast between the two devices. Device A exhibits a broader spectral wandering of 172~$\mu$eV for the emission line centered at 742.831~$\pm$~0.039~nm, which is calculated as $2\sigma$ where $\sigma$ is the standard deviation of the Gaussian. On the other hand, Device B shows a three-fold improvement with substantially reduced wandering down to 50~$\mu$eV centered at 749.071~$\pm$~0.006~nm. Moreover, the average linewidths also differ clearly, with Device A exhibiting 452~ $\pm$~58~$\mu$eV (0.201~$\pm$~0.026~nm) and Device B showing a significantly narrower linewidth of 132.5~ $\pm$~13.3~$\mu$eV (0.060~$\pm$0.006~nm), in good agreement with the single-shot spectra presented in Figs.~\ref{Figure2}A and B. These results demonstrate the importance of charge noise mitigation in these TMD single-photon emitters and provide important insights toward achieving transform-limited single-photon emission.

\begin{figure}[t!]
\includegraphics[width=0.8\columnwidth]{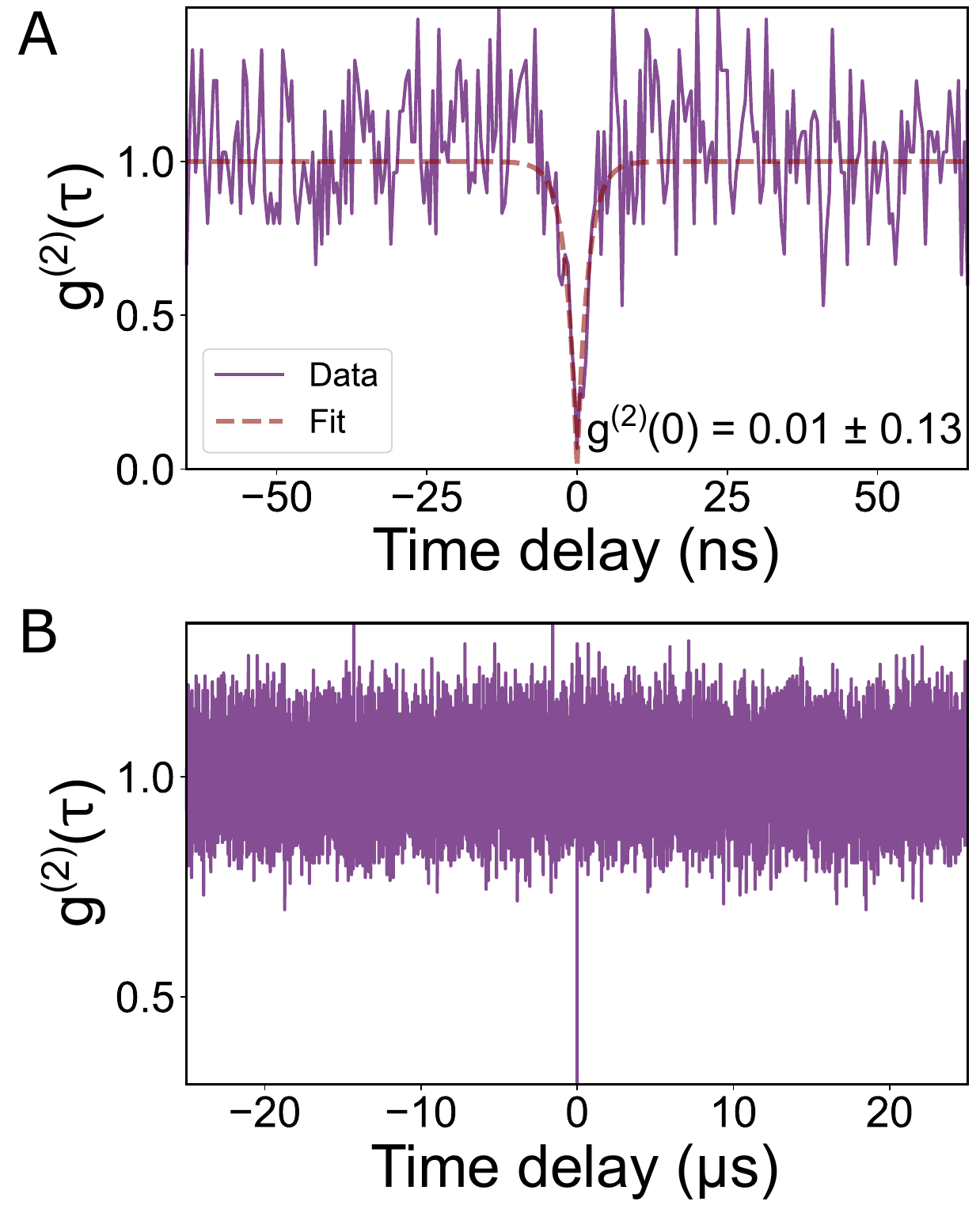}
\caption{\textbf{Second-order autocorrelation ($g^{(2)}(\tau)$) measurement under continuous-wave (CW) excitation for the exemplary emitter in Device B.} (\textbf{A}) Short time delay range ($\pm60$~ns) auto-correlation measurement. The measurement is normalized by dividing the coincidence counts by the calculated average over the extended time period. The fit (red dashed line) reveals a $g^{(2)}(0) = 0.01~\pm~0.13$, indicating a high single-photon purity of 99\% with the absence of any noticeable photon bunching. (\textbf{B}) The same auto-correlation measurement as in (A) on a longer time scale ($\pm25$~$\mu$s), showing no blinking or variations, demonstrating the emitter's exceptional photostability over extended periods.}
\label{Figure3}
\end{figure}

Next, to gain a deeper insight into the single-photon emission properties of the encapsulated emitter in Device B, we performed a second-order correlation measurement under 650 nm CW excitation.
Figure~\ref{Figure3}A shows the measured $g^{(2)}(\tau)$ over a short time window, revealing pronounced antibunching at zero delay with $g^{(2)}(0)~=~0.01~\pm~0.13$, confirming high single-photon purity of $\sim$99\%. Beyond the anti-bunching, $g^{(2)}(\tau)$ measurements also provide additional information about the emitter’s recombination dynamics.\cite{jahn2015artificial} For instance, the presence of photon bunching can indicate emission blinking, typically associated with substantial non-radiative recombination processes arising from an unstable charge environment, where fluctuating charges or trap states momentarily quench the emission.\cite{jahn2015artificial,zhai2020low} In the case of Device B, no bunching is observed, which may suggest stable emission dynamics without blinking on this short timescale. This observation is consistent with prior studies showing that the presence of bunching in WSe\textsubscript{2}-based quantum emitters is typically associated with the absence of hBN encapsulation, whereas its suppression is commonly observed in encapsulated devices.\cite{koperski2015single, tripathi2018spontaneous, Parto2021DefectK, daveau2020spectral} However, the second-order correlation measurements over shorter timescales do not fully capture potential blinking over longer durations. To assess this case, we extended the $g^{(2)}(\tau)$ measurement to a broader time window. As shown in Fig.~\ref{Figure3}B, the second-order correlation measurement $g^{(2)}(\tau \neq 0)$ remains flat over tens of $\mu$s, with no observable bunching. This behavior points to stable single-photon emission dynamics in Device B, with no evidence for blinking, even at longer timescales ($\pm$25~$\mu$s), highlighting the effectiveness of hBN encapsulation as a passive noise mitigation strategy.

% ------------------------------------------------------- %

\subsection{Active noise mitigation and tuning via electrical biasing}

\begin{figure}[htb!]
\includegraphics[width=1\columnwidth]{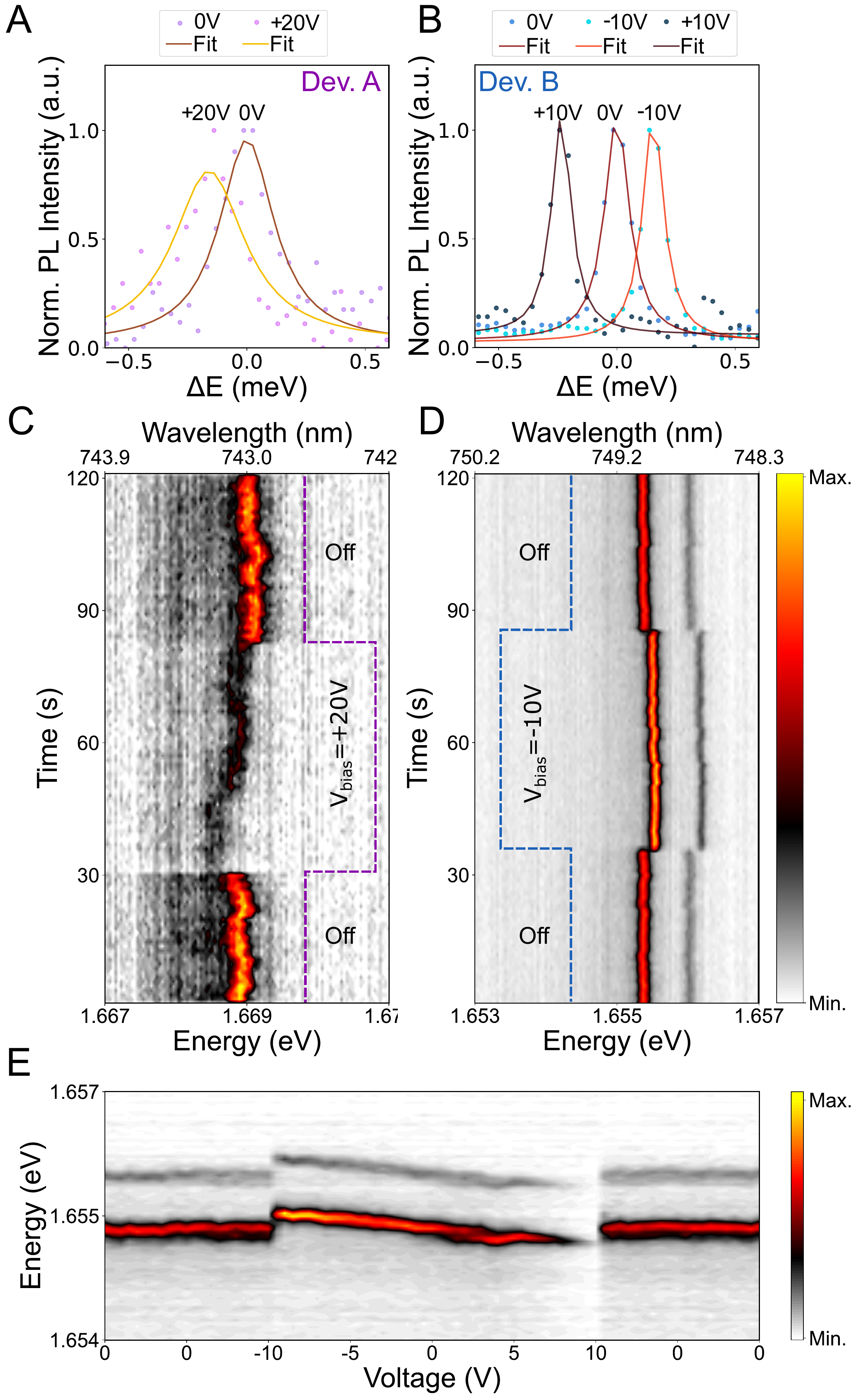}
\caption{\textbf{Bias effects on PL emission for unencapsulated (Dev. A) and encapsulated (Dev. B) WSe\textsubscript{2} QEs.} (\textbf{A,C}) High-resolution PL spectra with Lorentzian fitting under different biases. In Dev. A (A), a $+20$~V bias shifts the ZPL to lower energies while presenting a slight linewidth increase. Dev. B (C) shows a symmetric response: $-10$~V and $+10$~V shift the ZPL to higher and lower energies, respectively. (\textbf{B,D}) PL time traces under bias. Dev. A (B) exhibits an initial ZPL shift to lower energies, reverting to its original position with reduced intensity. Dev. B (D) shows a stable ZPL shift to higher energies with increased intensity under -10~V bias. (\textbf{E}) PL time trace for Dev. B under varying bias ($-10$~V to $+10$~V over 40~s). The ZPL shifts linearly (linear Stark effect), with the intensity dropping above $+7$~V. Data collected at 4~K under 650~nm CW laser excitation.}
\label{Figure4}
\end{figure}

Next, we investigate the role of electrical biasing in mitigating charge noise and modulating the emitter's optical properties.
Figure~\ref{Figure4}A presents $\mu$PL spectra ($T=4$~K, 650~nm CW laser excitation) from Device A, measured at $V_{bias}$~=~0~V and $V_{bias}$~=~$+20$~V. Upon applying a positive bias, we observe a Stark-induced redshift of the emission spectrum accompanied by noticeable spectral broadening. Interestingly, applying negative bias to Device A does not result in any noticeable change in the emission, likely due to an asymmetric local charge environment preferentially responding to one bias direction. To further investigate the effect of positive bias, we performed Lorentzian fits to the emission spectra at ~0~V and ~$+20$~V. The extracted results confirm our initial observations and show a clear spectral broadening with $W_\text{exp}$ increasing from $283~\pm~20$~$\mu$eV ($0.126~\pm~0.009$~nm) at 0~V to $370~\pm~39$~$\mu$eV ($0.165~\pm~0.018$~nm) at $+20$~V. Additionally, the fits also reveal a Stark-induced redshift of $\sim$100$~\mu$eV. 

To assess the temporal stability of this Stark shift effect, we recorded a PL time trace, shown in Figure~\ref{Figure4}C. Upon applying $V_{bias}$~=~$+20$~V at $t\sim 30$~s, the emission exhibits an immediate redshift; however, significant spectral fluctuations persist and the intensity drops sharply. This reduction may arise from decreased oscillator strength or reduced exciton capture efficiency, potentially due to carrier depletion or suppression of the local potential well under a positive bias. \cite{alen2007oscillator} Moreover, the emission line gradually drifts back toward its original position, despite the active biasing until $t\sim 85$~s. This behavior suggests a dynamically unstable charge environment that mimics capacitor-like charging and discharging, rather than maintaining a stable electric field, highlighting the limited effectiveness of electrical biasing in unencapsulated devices on a Si/SiO$_2$ substrate.

On the other hand, Device B exhibits a markedly different behavior under applied bias. Figure~\ref{Figure4}B presents normalized $\mu$PL spectra under $V_{bias}$~=~0~V, ~$+10$~V, and ~$-10$~V. In contrast to Device A, the encapsulated emitter responds symmetrically to both polarities, maintaining a narrow emission linewidth. Lorentzian fits yield $W_\text{exp}$~=~108~$\pm$~4~$\mu$eV ($0.049~\pm~0.002$~nm) at $V_{bias}$~=~$-10$~V and 101~$\pm$~6~$\mu$eV 
($0.045~\pm~0.003$~nm) at $V_{bias}$~=~$+10$~V, reaching the resolution limit of our setup. In comparison, at zero bias, the emitter displays a slightly broader linewidth of $125.07\pm4.14~\mu$eV ($0.057~\pm~0.002$~nm), in close agreement with the results of Fig.~\ref{Figure2}.

The emitter’s stability under negative bias is further examined in the $\mu$PL time-trace shown in Figure~\ref{Figure4}D. Upon applying $V_{bias}$~=~$-10$~V at $t\sim 30$~s, the emission peak undergoes an immediate blueshift of approximately 146~$\mu$eV and remains remarkably stable for the entire 50~s bias interval, before returning to its original position once the bias is removed. Notably, the emission intensity increases under bias, suggesting enhanced radiative decay, potentially due to the suppression of non-radiative decay channels. We employ the same fitting procedure used for the time-traces in Fig.~\ref{Figure2}, which reveals an average experimental linewidth of $W_\text{exp}$~=~104~$\pm$~5~$\mu$eV ($0.047~\pm~0.002$~nm), in close agreement with the snapshot presented in Fig.~\ref{Figure3}B. In addition, the fits yield a spectral wandering of 39~$\mu$eV, centered around an average emission wavelength of 749.005~$\pm$~0.008~nm. The time-trace in Figure~\ref{Figure4}D already shows resolution-limited linewidths, indicating strong suppression of spectral diffusion. To rigorously benchmark this performance, we performed high-resolution $\mu$PL measurements using a 1200 gr/mm grating and 0.25~s integration time, depicted in Supplementary Fig.~2A. The extracted parameters via Lorentzian fits reveal an average linewidth narrowing by $\sim$25\% under bias (Supplementary Fig.~2B), reaching the resolution limit at 101~$\pm$~6~$\mu$eV, in agreement with the value extracted from Fig. \ref{Figure4}B. Analysis of the central wavelength drift (Supplementary Fig.~2C) further confirms the long-term spectral stability, with spectral diffusion remaining constant at $\sim$40$~\mu$eV under bias. These results confirm that continuous electrical bias does not compromise long-timescale spectral stability while allowing the emitter to maintain resolution-limited emission.

In order to investigate the response of Device B under dynamic bias modulation, we recorded a PL time trace (Fig.~\ref{Figure4}E) during a linear voltage sweep from $V_{bias}$~=~$-10$~V to $+10$~V over 40~s (0.5~V step/frame), with an integration time of $\sim$0.8~s per frame. The emission energy shifts linearly and symmetrically with bias polarity, consistent with the linear Stark effect. However, the intensity drops sharply beyond $V_{bias}$~=~$+7$~V, and the emission is nearly quenched at $V_{bias}$~=~$+10$~V, which, similar to Device A, could be attributed to a possible excess charge imbalance activating non-radiative recombination channels, a reduction of the oscillator strength, or by inducing a transition between other excitonic states, such as from neutral to charged excitons (or vice versa). 
The quenching behaviour is studied further in a broader sweep range ($V_{bias}$=[$-20,+20$]~V) depicted in the Supplementary Fig.~3, with the rapid recovery, once the bias is removed, confirming the emitter's resilience.

To further assess the reproducibility of the tuning response, we analyzed additional emitters under both voltage ranges ($V_{bias}$~=~[$-10,+10$]~V and $V_{bias}$~=~[$-20,+20$]~V), as shown in the Supplementary Fig.~4, with exact quenching thresholds varying. Some emitters, such as Q3, exhibit distinct responses, likely due to differences in their initial excitonic states (neutral or positively/negatively charged states) or variations in the local electrostatic environment.

To quantitatively analyze the trends observed in the voltage sweep of Fig.~\ref{Figure4}E, we conducted a detailed analysis using the same fitting procedure as in Fig.~\ref{Figure4}D, the results of which are presented in the Supplementary Fig.~5. In Panel A, we observe a symmetric narrowing of the experimental linewidth ($W_\text{exp}$) with increasing bias magnitude, consistently reaching the spectrometer’s resolution limit near $V_{bias}$~=~$\pm$10~V. This behavior aligns with observations from both static bias snapshots and constant-bias time traces. Panel B shows the extracted emission energies as a function of bias, revealing a linear and symmetric Stark shift with maximum detuning around $\pm$140~$\mu$eV. In addition, Panel C presents TRPL measurements under various bias conditions, revealing lifetime bias-driven modulations of up to $\pm$40\%, indicative of field-induced changes in recombination dynamics through the modification of the local charge environment. These extracted lifetimes are used to calculate the radiative-limited linewidths ($W_\text{rad}$) in the following section, where we systematically evaluate the extent to which each noise mitigation strategy brings the emission from TMD quantum emitters towards the transform limit. 

% ------------------------------------------------------- %
\begin{figure*}[ht!]
\includegraphics[width=0.85\textwidth]{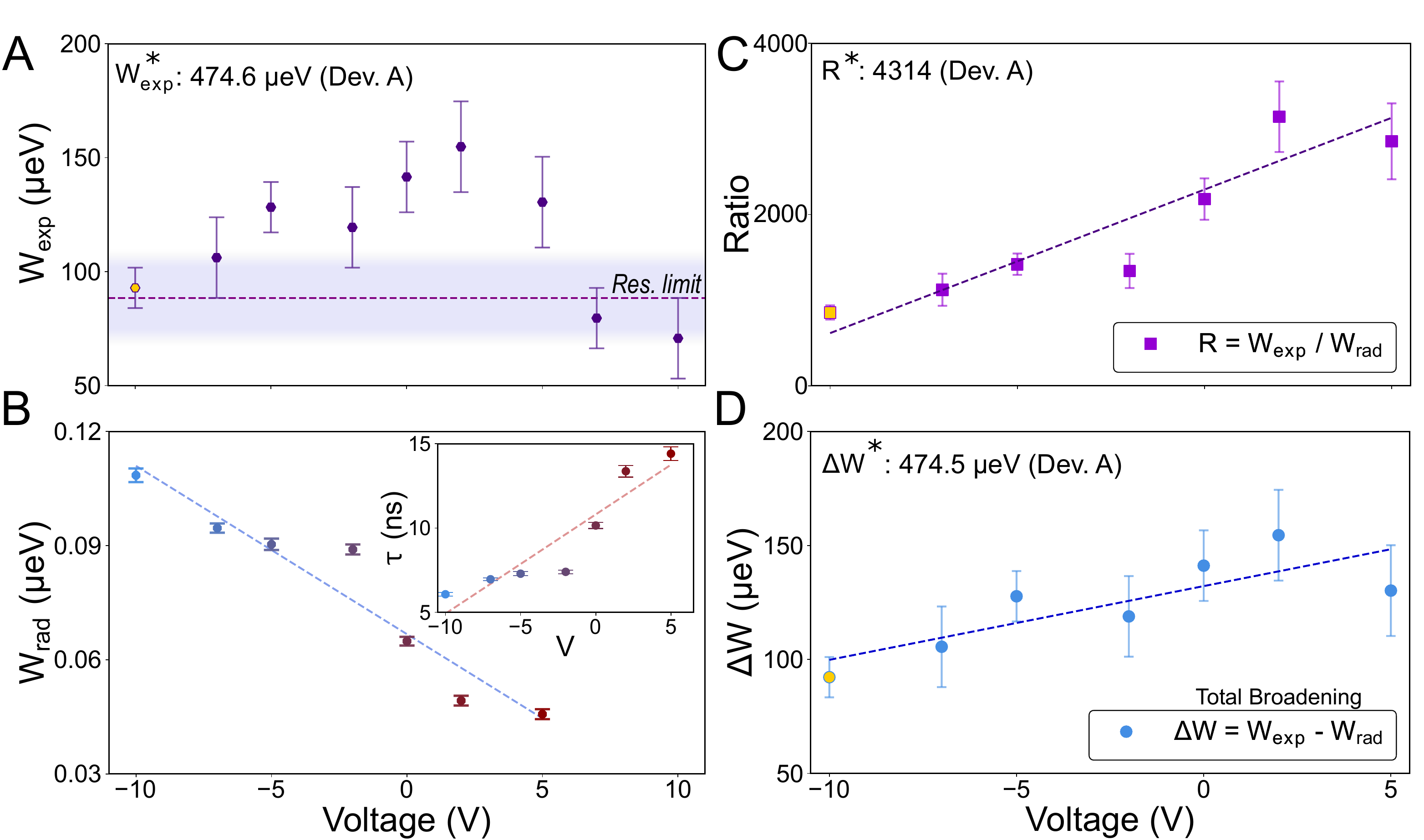}
\caption{\textbf{Voltage-dependent analysis of the measured ($W_{exp}$) and lifetime-limited ($W_\text{rad}$) linewidths and the correlated broadening mechanisms.} \textbf{(A)} Measured linewidth ($W_\text{exp}$) as a function of applied bias. At both extrema ($V_\text{bias}$~=~$\pm$10~V), $W_\text{exp}$ reaches the resolution limit  ($\sim$100~$\mu$\,eV), while near $V_\text{bias}$~=~0~V, it remains close to its intrinsic value. Notably, the data point at $V_\text{bias}$~=~$-10$~V is highlighted, as it corresponds to the highest emission intensity. In contrast, at $V_\text{bias}$~=~$+10$~V, the intensity is negligible, effectively rendering the emitter non-functional. Additionally, we include the value of the exemplary emitter in Device A at zero bias. This value is approximately three times larger than the encapsulated case at 0~V bias and more than five times larger than the values observed at $\pm$10~V. \textbf{(B)} Extracted lifetime-limited linewidths ($W_\text{rad}$) from time-resolved photoluminescence (TRPL) measurements as a function of applied bias. As the bias transitions from negative to positive values, the $W_\text{rad}$ decreases, indicating an increase in the emitter's lifetime. This inverse relationship is further illustrated in the inset, where the lifetime is plotted against bias, exhibiting the expected opposite trend, consistent with $W_\text{rad}\simeq$1/$\tau$. Additionally, the data points are color-coded to correspond to their respective bias values, for ease of interpretation. \textbf{(C)} The ratio $R = W_\text{exp} / W_\text{rad}$ as a function of voltage. The decreasing trend when moving towards negative voltage values arises from the simultaneous decrease of W$_\text{exp}$ and increase of W$_\text{rad}$, indicating that the two values approach each other in the voltage range where the emitter is most useful (i.e., higher emission intensity). The $R^{*}$ represents the ratio for Dev. A, twice larger than the unbiased case in Device B and five times larger than the optimized ($V_\text{bias}$~=~$-10$~V) case. \textbf{(D)} The total broadening of the emitter, defined as $\Delta W = W_\text{exp} - W_\text{rad}$, as a function of voltage. Once more, a decreasing trend is observed towards negative voltages, reaching below 100~$\mu$\,eV at $V_\text{bias}$~=~$-10$~V. The value for Dev. A ($\Delta W^{*}$) is more than 5 times larger than the lowest value achieved for Dev. B at $V_\text{bias}$~=~$-10$~V.}
\label{Figure5}
\end{figure*}

\subsection{Towards lifetime-limited PL emission}

Having demonstrated the effects of hBN encapsulation and electrical biasing on the optical properties of WSe\textsubscript{2} quantum emitters, we now assess how closely their single-photon emission has approached the transform-limited regime. To this end, we adopt two complementary figures of merit, inspired by methodologies in other solid-state systems. \cite{zhai2020low} The first is the linewidth ratio $R = W_\text{exp} / W_\text{rad}$, which compares the measured emission linewidth ($W_\text{exp}$) to the radiative-limited linewidth ($W_\text{rad}$), derived from TRPL measurements. This ratio provides a measure of how closely an emitter approaches the transform-limited regime, with $R = 1$ representing the ideal, noise-free case. The second figure of merit is the total broadening, $\Delta W = W_\text{exp} - W_\text{rad}$, which captures the absolute spectral broadening contributed by dephasing and noise-inducing mechanisms. We express the total measured linewidth as:
\begin{equation}
    W_\text{exp} = W_\text{rad} + W_\text{pd} + W_\text{noise}
    \label{eq:width_equation}
\end{equation}
where $W_\text{pd}$ accounts for broadening due to phonon dephasing, and $W_\text{noise}$ captures extrinsic fluctuations, primarily charge noise. While phonon coupling is an intrinsic limitation\cite{cadiz2017excitonic,chow2017unusual} and its mitigation falls outside the scope of this work, theoretical estimates suggest that $W_\text{noise}$ dominates the total broadening by up to four orders of magnitude, representing a key bottleneck for achieving indistinguishable single-photon emission in TMD quantum emitters.\cite{vannucci2024single} 

To systematically evaluate the impact of our noise mitigation strategies, we extract the experimental linewidths ($W_\text{exp}$) and radiative-limited linewidths ($W_\text{rad}$) as a function of applied biases from Device B. Device A, which serves as an unmitigated reference, is included only at zero bias. As summarized in Fig.~\ref{Figure5}A, hBN encapsulation alone reduces $W_\text{exp}$ from 474.6~$\mu$eV (Device A) to 141.5~$\mu$eV (Device B, $V_\text{bias} =$~0~V). With increasing bias, the linewidth narrows further, reaching below 100~$\mu$eV at $V_\text{bias} = \pm$10~V, limited by the spectrometer's resolution limit. Although a slightly narrower linewidth is observed for $V_\text{bias} =$~7 and 10~V, the value of $W_\text{exp}$ is less reliable since it is derived from data with a high signal-to-noise ratio near the resolution threshold.

Corresponding $W_\text{rad}$ values are calculated via $W_\text{rad} = \frac{h}{2\pi\tau}$,\cite{saleh2019fundamentals} assuming mono-exponential decay, based on the measured radiative lifetimes extracted from TRPL measurements at each bias voltage, as shown in Fig.~\ref{Figure5}B and its inset. The results reveal a clear bias-dependent trend, with lifetimes decreasing under negative bias and increasing under positive bias, exhibiting a total variation of approximately $\pm$40\% relative to the unbiased case. This variation translates into a corresponding shift in $W_\text{rad}$, with shorter lifetimes yielding broader radiative-limited linewidths. While $W_\text{rad}$ remains well below $W_\text{exp}$ under all conditions, the observed changes likely reflect a modulation in the emitter’s recombination dynamics due to perturbations to the local electrostatic environment induced by the applied bias. Because of significant PL quenching at higher bias, measurements beyond $V_\text{bias}~>$~$+5$~V are excluded.

With both $W_\text{exp}$ and $W_\text{rad}$ extracted across different bias conditions (Figs.~\ref{Figure5}A,B), we now combine these quantities to calculate the two key figures of merit that quantify the proximity to the transform-limited emission and the extent of charge-induced broadening. Figure~\ref{Figure5}C presents the linewidth ratio $R = W_\text{exp}/W_\text{rad}$, while Figure~\ref{Figure5}D shows the absolute broadening $\Delta W = W_\text{exp} - W_\text{rad}$, each compared against the unencapsulated Device A as a reference. 
We find that hBN encapsulation alone reduces $R$ significantly from $\sim$4314 in Device A to approximately $\sim$2100 in the unbiased Device B. Introducing electrical bias further suppresses $R$, reaching a minimum value of $\sim$860 at $V_\text{bias}$~=~$-10$~V. This nearly five-fold reduction brings the emitter substantially closer to the transform-limited regime. It is worth noting, however, that $R$ depends strongly on the radiative lifetime, which can vary widely among TMD emitters, from sub-nanosecond to tens of nanoseconds. Consequently, values of $R$ calculated based on the results presented in other works span a broad range, from below 100 to several thousand (see Supplementary Table 1). In our case, despite $W_\text{exp}$ approaching the resolution limit, the minimum $R$ remains ~860—higher than the best-reported value of $R\sim270$ for WSe\textsubscript{2} emitters with lifetimes around 1.5 ns.\cite{srivastava2015optically} This apparent discrepancy can be misleading when evaluating proximity to the transform-limited regime, and is caused solely by the difference in radiative lifetimes, well inside the expected range. 

The absolute spectral broadening $\Delta W$, shown in Fig.~\ref{Figure5}D, follows a similar trend. Encapsulation alone reduces $\Delta W$ from 474.5~$\mu$eV (Device A) to $\sim$140~$\mu$eV (Device B), while electrical bias further narrows it to a minimum of 92.8~$\mu$eV at $V_\text{bias}$~=~$-10$~V. This more than five-fold reduction confirms that combining hBN encapsulation with electrical biasing is highly effective in suppressing charge noise and subsequently narrowing spectral linewidths in WSe\textsubscript{2} quantum emitters. Notably, the value of $\Delta W$ at $V_\text{bias}$~=~$-10$~V surpasses the best previously reported for WSe\textsubscript{2} and related materials, indicating that such levels of noise suppression can be reached without relying on complex fabrication strategies or cavity-enhanced emission.\cite{kumar2015strain} Still, as with $R$, care must be taken in interpreting $\Delta W$, since it is fundamentally limited by the spectral resolution used to measure $W_\text{exp}$. Many studies lack the high-resolution techniques necessary to resolve intrinsic linewidths, and instead report lower-bound estimates that sacrifice resolution in favor of other experimental priorities. In contrast, our measurements already reach the resolution limit of a standard optical spectrometer, indicating that the true intrinsic linewidth may be even narrower. This underscores the need for advanced spectroscopic approaches, such as interferometric or cavity-enhanced techniques, to accurately resolve the transform-limited emission regime in future studies. 

Beyond demonstrating the effectiveness of hBN encapsulation and electrical biasing in suppressing charge noise, our analysis of both $R$ and $\Delta W$ provides a practical and quantitative framework for benchmarking the performance of quantum emitters in TMD materials. Together, these complementary figures of merit enable a more nuanced and reliable assessment of emitter quality and proximity to the transform-limited regime.

\section{Discussion }

Our findings demonstrate that the combination of hBN encapsulation and electrostatic biasing offers an effective strategy for mitigating charge noise in WSe\textsubscript{2} quantum emitters, advancing their performance toward the transform limit of single-photon emission. Encapsulation alone leads to substantial improvements in single-photon emission, evidenced by significantly reduced linewidths and suppressed spectral wandering. These enhancements are attributed to the electrostatic screening provided by the hBN layers, which isolate the emitter from a noisy charge environment (e.g. substrate). 
Supporting this interpretation, TRPL measurements reveal a transition from biexponential decay, indicative of non-radiative decay channels, in unencapsulated emitters, to mono-exponential decay in encapsulated devices. Furthermore, second-order correlation measurements confirm high single-photon purity of 99\%, with no observable bunching even at longer timescales ($\pm25~\mu$s), suggesting the absence of blinking in the device.

Electrostatic biasing provides an additional enhancement in emitter performance while allowing for tunability and control. However, its benefits are only realized when used in conjunction with encapsulation. While unencapsulated devices suffer from increased instability and possible quenching under bias, encapsulated emitters exhibit robust and reversible linear Stark tuning and further linewidth narrowing. Remarkably, the linewidths under bias reach the resolution limit of our spectrometer, without compromising long-term spectral stability. Simultaneously, we observe a radiative lifetime modulation of up to $\pm40\%$, suggesting that biasing can also influence the recombination dynamics in the system.

To quantify the impact of our noise mitigation strategies, we introduced two figures of merit: the linewidth ratio $R = W_\text{exp} / W_\text{rad}$ and the total broadening $\Delta W = W_\text{exp} - W_\text{rad}$. These metrics provide a consistent and quantitative measure of how close an emitter approaches transform-limited behavior. In Device B, we observe a fivefold reduction in both $R$ (from 4314 to 860) and $\Delta W$ (from 474.5 to 92.8~$\mu\text{eV}$) relative to the unencapsulated and unbiased reference (Device A). While the measured $W_\text{exp}$ reaches our setup's resolution limit, it does not reflect the true intrinsic linewidth of the emitter. This highlights the need for higher-resolution spectroscopic techniques to resolve this regime. Moreover, the relatively high value of $R$, driven by the long radiative lifetime ($\sim$6~ns), likely underestimates the actual proximity to the transform limit.

To that end, there are several viable pathways to reduce $R$ even closer to the transform limit. A promising approach lies in implementing advanced optical excitation schemes, e.g., phonon-assisted, resonant, and SUPER, in contrast to a standard above-band excitation.\cite{karli2022super, kumar2016resonant,piccinini2024high, boos2024coherent} For instance, phonon-assisted excitation schemes have been shown to reduce radiative lifetimes to $\sim$1 ns for the case of WSe\textsubscript{2} quantum emitters.\cite{piccinini2024high} If implemented in our Device B under a proper bias, such an emitter with reduced lifetime could yield an $R$ value of $\sim$150, bringing it significantly closer to the transform limit. Furthermore, integrating our Device B into an optical cavity with a modest Purcell factor of $\sim$10, combined with an optimized excitation scheme, could reduce the lifetime to $\sim$0.1~ns.\cite{iff2021purcell, azzam2023purcell, drawer2023monolayer} This would enable an improved $R$ value of $\sim$15, just one order of magnitude away from the ideal transform-limit performance. 

\section{Conclusion}

In conclusion, we demonstrated how the combination of hBN encapsulation and electrostatic biasing can improve the emission properties of WSe\textsubscript{2} quantum emitters, while also providing a consistent framework for benchmarking their performance. Specifically, we showed that hBN encapsulation alone reduces both the emission linewidth and spectral wandering, bringing them down to approximately 150~$\mu$eV and 40~$\mu$eV, respectively. In addition, time-resolved measurements confirm a stable electrostatic environment through a mono-exponential decay, and second-order correlation measurements reveal a high single-photon purity ($\sim$99\%) with no observable blinking. When electrostatic biasing is applied in conjunction with encapsulation, the emitter performance is further enhanced. The linewidth reaches the resolution limit ($\sim100~\mu$eV), without compromising long-term stability, and the emission energy becomes tunable over a range exceeding 280~$\mu$eV. The progress towards transform-limited emission is assessed through two figures of merit, the linewidth ratio ($R$) and the total broadening ($\Delta W$), both of which show more than a five-fold improvement in the optimized devices. These results provide a clear pathway toward the development of low-noise and tunable WSe\textsubscript{2} quantum emitters, potentially realizing electrically controllable sources of indistinguishable single-photons for future photonic quantum technologies.

\section{Materials and Methods} \label{section:methods}

\subsection{Sample preparation}

\textit{Nanopillar fabrication:} The wafer consists of 110~nm of SiO$_2$ grown by thermal oxidation on a Si wafer. Alignment marks were fabricated by spin coating and patterning UV resist (AZ 5214-E, MicroChemicals), followed by physical vapor deposition of 5/50~nm of Ti/Au and a liftoff process. 
For the fabrication of the nanopillars, a small chip is cleaved from the wafer and spin-coated with a high-resolution negative e-beam resist (hydrogen silsesquioxane or HSQ, XR-1541-006, Dow Corning) at 3000~rpm for 1~min, followed by two soft baking steps at 120~$^\circ$C and 220~$^\circ$C, both for 2 minutes.
The resist was patterned (JBX-9500FS, JEOL, 100 kV, 6 nA) with a dose of 11000 \textmu C/cm$^2$ and developed in a 1:3 solution of AZ 400K : H$_2$O to obtain 150~nm- (Device A) or 250~nm-tall (Device B) nanopillars in the shape of a three-pointed star. 

\textit{Exfoliation and transferring:} WSe\textsubscript{2} flakes were first exfoliated from bulk crystals (HQ Graphene) via the scotch-tape method \cite{castellanos2014deterministic}. The flakes were then deposited on a polydimethylsiloxane (PDMS) stamp, which was prepared on a glass slide. The monolayers were identified by photoluminescence measurements using an optical microscope (Nikon 20×, NA = 0.45) and a 450~nm LED excitation source. The transfer of the flakes to the nanostructures was performed with the help of a transfer stage that could heat the chip to 70~$^\circ$C.
At this temperature, the van der Waals interactions between the flake and the substrate overcome the adhesion to PDMS, and the flake is released from the polymer.

\textit{Defect fabrication:} Defects were introduced on the monolayer lattice by bombarding the strained areas of the material with an electron beam (JBX-9500FS, JEOL, 100 kV, 1000 \textmu C/cm$^2$).

\subsection{Optical characterization} 

\textit{Photoluminescence measurements:} The PL measurements were obtained with a custom-built low-temperature micro-photoluminescence (\textmu PL) setup. The sample is mounted on a closed-cycle cryostat (attoDRY800, Attocube) operating at a base temperature of 4 K. The cryostat is equipped with piezoelectric nanopositioners and a low-temperature microscope objective ($60 \times$, NA = 0.82) located inside the cryostat. For the \textmu PL spectroscopy, a CW and pulsed (20/80 MHz) laser diode at 650~nm (LDH-D-C-650, PicoQuant) was employed for the optical excitation of the emitters. The PL spectra were acquired with a fiber-coupled signal directed to a spectrometer (iHR 550, Horiba) equipped with 600 or 1200 lines per mm gratings and a CCD camera (Syncerity, Horiba). 

\textit{Time-correlated measurements:} Time-correlated single-photon-counting mode was used to perform time-resolved PL characterization and single-photon statistics measurements. For second-order correlation measurements, we employed a fiber-optic-based Hanbury-Brown and Twiss (HBT) interferometer. The photon counting was conducted using superconducting nanowires single-photon detectors (ID218, ID Quantique) connected to a time-correlated single-photon counting module (Time Tagger Ultra, Swabian Instruments).

\section*{Acknowledgements}
The authors acknowledge the cleanroom facilities at DTU Nanolab – National Centre for Nano Fabrication and Characterization. 

\section*{Funding}

The authors acknowledge support from the European Research Council (ERC-StG ``TuneTMD", grant no. 101076437); 
the Villum Foundation (grant no. VIL53033). 
The authors also acknowledge the European Research Council (ERC-CoG ``Unity", grant no. 865230); the TICRA foundation;
and the Carlsberg Foundation (grant no. CF21-0496). 

%\section*{Author Contributions}

\section*{Competing Interests}
The authors declare no competing financial interest.

\section*{Data and materials availability}
\noindent
All data needed to verify the results and conclusions of this publication are present in the main paper and the Supplementary Materials.

\section{References}
\bibliography{ref} % refreshed manually

\end{document}

%% file: settings.tex
\usepackage[intlimits]{amsmath}
\usepackage{amssymb}
\usepackage{amsthm}
\usepackage{booktabs}
\usepackage{braket}
\usepackage{color}
\usepackage{float}
\usepackage{graphicx}
\usepackage{glossaries}
\usepackage[colorlinks=true,
            linkcolor=black,
            citecolor=black,
            filecolor=black,
            urlcolor=black
            ]{hyperref}
\usepackage[utf8]{inputenc}
\usepackage[version=4]{mhchem}
\usepackage{multirow}
\usepackage{soul}
\usepackage{tabularx}
\usepackage{units}
\usepackage[dvipsnames]{xcolor}

\usepackage{dcolumn}
\usepackage{float}
\usepackage{booktabs}

\setacronymstyle{long-short}
\newacronym{qd}{QD}{quantum dot}
\newacronym{pl}{PL}{Photoluminescence}
\newacronym{cl}{CL}{Cathodoluminescence}
\newacronym{gaas}{GaAs}{Gallium Arsenide}
\newacronym{ingaas}{InGaAs}{Indium Gallium Arsenide}
\newacronym{ebl}{EBL}{Electron beam lithography}

\newacronym{vis}{VIS}{visible}
\newacronym{nir}{NIR}{near-infrared}
\newacronym{uv}{UV}{ultraviolet}
\newacronym{mse}{MSE}{mean squared error}
\newacronym{si}{SI}{Supporting Information}

%% file: ms.bbl
\begin{thebibliography}{60}%
\makeatletter
\providecommand \@ifxundefined [1]{%
 \@ifx{#1\undefined}
}%
\providecommand \@ifnum [1]{%
 \ifnum #1\expandafter \@firstoftwo
 \else \expandafter \@secondoftwo
 \fi
}%
\providecommand \@ifx [1]{%
 \ifx #1\expandafter \@firstoftwo
 \else \expandafter \@secondoftwo
 \fi
}%
\providecommand \natexlab [1]{#1}%
\providecommand \enquote  [1]{``#1''}%
\providecommand \bibnamefont  [1]{#1}%
\providecommand \bibfnamefont [1]{#1}%
\providecommand \citenamefont [1]{#1}%
\providecommand \href@noop [0]{\@secondoftwo}%
\providecommand \href [0]{\begingroup \@sanitize@url \@href}%
\providecommand \@href[1]{\@@startlink{#1}\@@href}%
\providecommand \@@href[1]{\endgroup#1\@@endlink}%
\providecommand \@sanitize@url [0]{\catcode `\\12\catcode `\$12\catcode `\&12\catcode `\#12\catcode `\^12\catcode `\_12\catcode `\%12\relax}%
\providecommand \@@startlink[1]{}%
\providecommand \@@endlink[0]{}%
\providecommand \url  [0]{\begingroup\@sanitize@url \@url }%
\providecommand \@url [1]{\endgroup\@href {#1}{\urlprefix }}%
\providecommand \urlprefix  [0]{URL }%
\providecommand \Eprint [0]{\href }%
\providecommand \doibase [0]{http://dx.doi.org/}%
\providecommand \selectlanguage [0]{\@gobble}%
\providecommand \bibinfo  [0]{\@secondoftwo}%
\providecommand \bibfield  [0]{\@secondoftwo}%
\providecommand \translation [1]{[#1]}%
\providecommand \BibitemOpen [0]{}%
\providecommand \bibitemStop [0]{}%
\providecommand \bibitemNoStop [0]{.\EOS\space}%
\providecommand \EOS [0]{\spacefactor3000\relax}%
\providecommand \BibitemShut  [1]{\csname bibitem#1\endcsname}%
\let\auto@bib@innerbib\@empty
%</preamble>
\bibitem [{\citenamefont {Gregersen}\ \emph {et~al.}(2017)\citenamefont {Gregersen}, \citenamefont {McCutcheon},\ and\ \citenamefont {Mørk}}]{gregersen_mørk_2017}%
  \BibitemOpen
  \bibfield  {author} {\bibinfo {author} {\bibfnamefont {Niels}\ \bibnamefont {Gregersen}}, \bibinfo {author} {\bibfnamefont {Dara~P.S.}\ \bibnamefont {McCutcheon}}, \ and\ \bibinfo {author} {\bibfnamefont {Jesper}\ \bibnamefont {Mørk}},\ }\enquote {\bibinfo {title} {Single-photon sources},}\ in\ \href {\doibase https://doi.org/10.4324/9781315152318} {\emph {\bibinfo {booktitle} {Handbook of Optoelectronic Device Modeling and Simulation}}},\ Vol.~\bibinfo {volume} {2}\ (\bibinfo  {publisher} {CRC Press},\ \bibinfo {year} {2017})\ p.\ \bibinfo {pages} {585–604},\ \bibinfo {edition} {1st}\ ed.\BibitemShut {Stop}%
\bibitem [{\citenamefont {Gao}\ \emph {et~al.}(2012)\citenamefont {Gao}, \citenamefont {Fallahi}, \citenamefont {Togan}, \citenamefont {Miguel-S{\'a}nchez},\ and\ \citenamefont {Imamoglu}}]{gao2012observation}%
  \BibitemOpen
  \bibfield  {author} {\bibinfo {author} {\bibfnamefont {W.B.}\ \bibnamefont {Gao}}, \bibinfo {author} {\bibfnamefont {Parisa}\ \bibnamefont {Fallahi}}, \bibinfo {author} {\bibfnamefont {Emre}\ \bibnamefont {Togan}}, \bibinfo {author} {\bibfnamefont {Javier}\ \bibnamefont {Miguel-S{\'a}nchez}}, \ and\ \bibinfo {author} {\bibfnamefont {Atac}\ \bibnamefont {Imamoglu}},\ }\bibfield  {title} {\enquote {\bibinfo {title} {Observation of entanglement between a quantum dot spin and a single photon},}\ }\href@noop {} {\bibfield  {journal} {\bibinfo  {journal} {Nature}\ }\textbf {\bibinfo {volume} {491}},\ \bibinfo {pages} {426--430} (\bibinfo {year} {2012})}\BibitemShut {NoStop}%
\bibitem [{\citenamefont {Wang}\ \emph {et~al.}(2020)\citenamefont {Wang}, \citenamefont {Sciarrino}, \citenamefont {Laing},\ and\ \citenamefont {Thompson}}]{wang2020integrated}%
  \BibitemOpen
  \bibfield  {author} {\bibinfo {author} {\bibfnamefont {Jianwei}\ \bibnamefont {Wang}}, \bibinfo {author} {\bibfnamefont {Fabio}\ \bibnamefont {Sciarrino}}, \bibinfo {author} {\bibfnamefont {Anthony}\ \bibnamefont {Laing}}, \ and\ \bibinfo {author} {\bibfnamefont {Mark~G}\ \bibnamefont {Thompson}},\ }\bibfield  {title} {\enquote {\bibinfo {title} {Integrated photonic quantum technologies},}\ }\href@noop {} {\bibfield  {journal} {\bibinfo  {journal} {Nature Photonics}\ }\textbf {\bibinfo {volume} {14}},\ \bibinfo {pages} {273--284} (\bibinfo {year} {2020})}\BibitemShut {NoStop}%
\bibitem [{\citenamefont {Xu}\ \emph {et~al.}(2020)\citenamefont {Xu}, \citenamefont {Ma}, \citenamefont {Zhang}, \citenamefont {Lo},\ and\ \citenamefont {Pan}}]{xu2020secure}%
  \BibitemOpen
  \bibfield  {author} {\bibinfo {author} {\bibfnamefont {Feihu}\ \bibnamefont {Xu}}, \bibinfo {author} {\bibfnamefont {Xiongfeng}\ \bibnamefont {Ma}}, \bibinfo {author} {\bibfnamefont {Qiang}\ \bibnamefont {Zhang}}, \bibinfo {author} {\bibfnamefont {Hoi-Kwong}\ \bibnamefont {Lo}}, \ and\ \bibinfo {author} {\bibfnamefont {Jian-Wei}\ \bibnamefont {Pan}},\ }\bibfield  {title} {\enquote {\bibinfo {title} {Secure quantum key distribution with realistic devices},}\ }\href@noop {} {\bibfield  {journal} {\bibinfo  {journal} {Reviews of Modern Physics}\ }\textbf {\bibinfo {volume} {92}},\ \bibinfo {pages} {025002} (\bibinfo {year} {2020})}\BibitemShut {NoStop}%
\bibitem [{\citenamefont {Senellart}\ \emph {et~al.}(2017)\citenamefont {Senellart}, \citenamefont {Solomon},\ and\ \citenamefont {White}}]{senellart2017high}%
  \BibitemOpen
  \bibfield  {author} {\bibinfo {author} {\bibfnamefont {Pascale}\ \bibnamefont {Senellart}}, \bibinfo {author} {\bibfnamefont {Glenn}\ \bibnamefont {Solomon}}, \ and\ \bibinfo {author} {\bibfnamefont {Andrew}\ \bibnamefont {White}},\ }\bibfield  {title} {\enquote {\bibinfo {title} {High-performance semiconductor quantum-dot single-photon sources},}\ }\href@noop {} {\bibfield  {journal} {\bibinfo  {journal} {Nature Nanotechnology}\ }\textbf {\bibinfo {volume} {12}},\ \bibinfo {pages} {1026--1039} (\bibinfo {year} {2017})}\BibitemShut {NoStop}%
\bibitem [{\citenamefont {Ding}\ \emph {et~al.}(2016)\citenamefont {Ding}, \citenamefont {He}, \citenamefont {Duan}, \citenamefont {Gregersen}, \citenamefont {Chen}, \citenamefont {Unsleber}, \citenamefont {Maier}, \citenamefont {Schneider}, \citenamefont {Kamp}, \citenamefont {H{\"o}fling} \emph {et~al.}}]{ding2016demand}%
  \BibitemOpen
  \bibfield  {author} {\bibinfo {author} {\bibfnamefont {Xing}\ \bibnamefont {Ding}}, \bibinfo {author} {\bibfnamefont {Yu}~\bibnamefont {He}}, \bibinfo {author} {\bibfnamefont {Z-C}\ \bibnamefont {Duan}}, \bibinfo {author} {\bibfnamefont {Niels}\ \bibnamefont {Gregersen}}, \bibinfo {author} {\bibfnamefont {M-C}\ \bibnamefont {Chen}}, \bibinfo {author} {\bibfnamefont {S}~\bibnamefont {Unsleber}}, \bibinfo {author} {\bibfnamefont {Sebastian}\ \bibnamefont {Maier}}, \bibinfo {author} {\bibfnamefont {Christian}\ \bibnamefont {Schneider}}, \bibinfo {author} {\bibfnamefont {Martin}\ \bibnamefont {Kamp}}, \bibinfo {author} {\bibfnamefont {Sven}\ \bibnamefont {H{\"o}fling}},  \emph {et~al.},\ }\bibfield  {title} {\enquote {\bibinfo {title} {On-demand single photons with high extraction efficiency and near-unity indistinguishability from a resonantly driven quantum dot in a micropillar},}\ }\href@noop {} {\bibfield  {journal} {\bibinfo  {journal} {Physical Review Letters}\ }\textbf {\bibinfo {volume} {116}},\
  \bibinfo {pages} {020401} (\bibinfo {year} {2016})}\BibitemShut {NoStop}%
\bibitem [{\citenamefont {Thomas}\ \emph {et~al.}(2021)\citenamefont {Thomas}, \citenamefont {Billard}, \citenamefont {Coste}, \citenamefont {Wein}, \citenamefont {Priya}, \citenamefont {Ollivier}, \citenamefont {Krebs}, \citenamefont {Taza{\"\i}rt}, \citenamefont {Harouri}, \citenamefont {Lemaitre} \emph {et~al.}}]{thomas2021bright}%
  \BibitemOpen
  \bibfield  {author} {\bibinfo {author} {\bibfnamefont {S.E.}\ \bibnamefont {Thomas}}, \bibinfo {author} {\bibfnamefont {M.}~\bibnamefont {Billard}}, \bibinfo {author} {\bibfnamefont {N.}~\bibnamefont {Coste}}, \bibinfo {author} {\bibfnamefont {S.C.}\ \bibnamefont {Wein}}, \bibinfo {author} {\bibnamefont {Priya}}, \bibinfo {author} {\bibfnamefont {H.}~\bibnamefont {Ollivier}}, \bibinfo {author} {\bibfnamefont {Olivier}\ \bibnamefont {Krebs}}, \bibinfo {author} {\bibfnamefont {L.}~\bibnamefont {Taza{\"\i}rt}}, \bibinfo {author} {\bibfnamefont {A.}~\bibnamefont {Harouri}}, \bibinfo {author} {\bibfnamefont {A.}~\bibnamefont {Lemaitre}},  \emph {et~al.},\ }\bibfield  {title} {\enquote {\bibinfo {title} {Bright polarized single-photon source based on a linear dipole},}\ }\href@noop {} {\bibfield  {journal} {\bibinfo  {journal} {Physical Review Letters}\ }\textbf {\bibinfo {volume} {126}},\ \bibinfo {pages} {233601} (\bibinfo {year} {2021})}\BibitemShut {NoStop}%
\bibitem [{\citenamefont {Karli}\ \emph {et~al.}(2024)\citenamefont {Karli}, \citenamefont {Vajner}, \citenamefont {Kappe}, \citenamefont {Hagen}, \citenamefont {Hansen}, \citenamefont {Schwarz}, \citenamefont {Bracht}, \citenamefont {Schimpf}, \citenamefont {Covre~da Silva}, \citenamefont {Walther} \emph {et~al.}}]{karli2024controlling}%
  \BibitemOpen
  \bibfield  {author} {\bibinfo {author} {\bibfnamefont {Yusuf}\ \bibnamefont {Karli}}, \bibinfo {author} {\bibfnamefont {Daniel~A.}\ \bibnamefont {Vajner}}, \bibinfo {author} {\bibfnamefont {Florian}\ \bibnamefont {Kappe}}, \bibinfo {author} {\bibfnamefont {Paul~C.A.}\ \bibnamefont {Hagen}}, \bibinfo {author} {\bibfnamefont {Lena~M.}\ \bibnamefont {Hansen}}, \bibinfo {author} {\bibfnamefont {Ren{\'e}}\ \bibnamefont {Schwarz}}, \bibinfo {author} {\bibfnamefont {Thomas~K.}\ \bibnamefont {Bracht}}, \bibinfo {author} {\bibfnamefont {Christian}\ \bibnamefont {Schimpf}}, \bibinfo {author} {\bibfnamefont {Saimon~F.}\ \bibnamefont {Covre~da Silva}}, \bibinfo {author} {\bibfnamefont {Philip}\ \bibnamefont {Walther}},  \emph {et~al.},\ }\bibfield  {title} {\enquote {\bibinfo {title} {Controlling the photon number coherence of solid-state quantum light sources for quantum cryptography},}\ }\href@noop {} {\bibfield  {journal} {\bibinfo  {journal} {npj Quantum Information}\ }\textbf {\bibinfo {volume} {10}},\ \bibinfo
  {pages} {17} (\bibinfo {year} {2024})}\BibitemShut {NoStop}%
\bibitem [{\citenamefont {Boos}\ \emph {et~al.}(2024)\citenamefont {Boos}, \citenamefont {Sbresny}, \citenamefont {Kim}, \citenamefont {Kremser}, \citenamefont {Riedl}, \citenamefont {Bopp}, \citenamefont {Rauhaus}, \citenamefont {Scaparra}, \citenamefont {J{\"o}ns}, \citenamefont {Finley} \emph {et~al.}}]{boos2024coherent}%
  \BibitemOpen
  \bibfield  {author} {\bibinfo {author} {\bibfnamefont {Katarina}\ \bibnamefont {Boos}}, \bibinfo {author} {\bibfnamefont {Friedrich}\ \bibnamefont {Sbresny}}, \bibinfo {author} {\bibfnamefont {Sang~Kyu}\ \bibnamefont {Kim}}, \bibinfo {author} {\bibfnamefont {Malte}\ \bibnamefont {Kremser}}, \bibinfo {author} {\bibfnamefont {Hubert}\ \bibnamefont {Riedl}}, \bibinfo {author} {\bibfnamefont {Frederik~W}\ \bibnamefont {Bopp}}, \bibinfo {author} {\bibfnamefont {William}\ \bibnamefont {Rauhaus}}, \bibinfo {author} {\bibfnamefont {Bianca}\ \bibnamefont {Scaparra}}, \bibinfo {author} {\bibfnamefont {Klaus~D}\ \bibnamefont {J{\"o}ns}}, \bibinfo {author} {\bibfnamefont {Jonathan~J}\ \bibnamefont {Finley}},  \emph {et~al.},\ }\bibfield  {title} {\enquote {\bibinfo {title} {Coherent swing-up excitation for semiconductor quantum dots},}\ }\href@noop {} {\bibfield  {journal} {\bibinfo  {journal} {Advanced Quantum Technologies}\ }\textbf {\bibinfo {volume} {7}},\ \bibinfo {pages} {2300359} (\bibinfo {year}
  {2024})}\BibitemShut {NoStop}%
\bibitem [{\citenamefont {Wei}\ \emph {et~al.}(2022)\citenamefont {Wei}, \citenamefont {Liu}, \citenamefont {Li}, \citenamefont {Yu}, \citenamefont {Su}, \citenamefont {Li}, \citenamefont {Shang}, \citenamefont {Liu}, \citenamefont {Hao}, \citenamefont {Ni} \emph {et~al.}}]{wei2022tailoring}%
  \BibitemOpen
  \bibfield  {author} {\bibinfo {author} {\bibfnamefont {Yuming}\ \bibnamefont {Wei}}, \bibinfo {author} {\bibfnamefont {Shunfa}\ \bibnamefont {Liu}}, \bibinfo {author} {\bibfnamefont {Xueshi}\ \bibnamefont {Li}}, \bibinfo {author} {\bibfnamefont {Ying}\ \bibnamefont {Yu}}, \bibinfo {author} {\bibfnamefont {Xiangbin}\ \bibnamefont {Su}}, \bibinfo {author} {\bibfnamefont {Shulun}\ \bibnamefont {Li}}, \bibinfo {author} {\bibfnamefont {Xiangjun}\ \bibnamefont {Shang}}, \bibinfo {author} {\bibfnamefont {Hanqing}\ \bibnamefont {Liu}}, \bibinfo {author} {\bibfnamefont {Huiming}\ \bibnamefont {Hao}}, \bibinfo {author} {\bibfnamefont {Haiqiao}\ \bibnamefont {Ni}},  \emph {et~al.},\ }\bibfield  {title} {\enquote {\bibinfo {title} {Tailoring solid-state single-photon sources with stimulated emissions},}\ }\href@noop {} {\bibfield  {journal} {\bibinfo  {journal} {Nature Nanotechnology}\ }\textbf {\bibinfo {volume} {17}},\ \bibinfo {pages} {470--476} (\bibinfo {year} {2022})}\BibitemShut {NoStop}%
\bibitem [{\citenamefont {Montblanch}\ \emph {et~al.}(2023)\citenamefont {Montblanch}, \citenamefont {Barbone}, \citenamefont {Aharonovich}, \citenamefont {Atat{\"u}re},\ and\ \citenamefont {Ferrari}}]{montblanch2023layered}%
  \BibitemOpen
  \bibfield  {author} {\bibinfo {author} {\bibfnamefont {Alejandro R.-P.}\ \bibnamefont {Montblanch}}, \bibinfo {author} {\bibfnamefont {Matteo}\ \bibnamefont {Barbone}}, \bibinfo {author} {\bibfnamefont {Igor}\ \bibnamefont {Aharonovich}}, \bibinfo {author} {\bibfnamefont {Mete}\ \bibnamefont {Atat{\"u}re}}, \ and\ \bibinfo {author} {\bibfnamefont {Andrea~C.}\ \bibnamefont {Ferrari}},\ }\bibfield  {title} {\enquote {\bibinfo {title} {Layered materials as a platform for quantum technologies},}\ }\href@noop {} {\bibfield  {journal} {\bibinfo  {journal} {Nature Nanotechnology}\ }\textbf {\bibinfo {volume} {18}},\ \bibinfo {pages} {555--571} (\bibinfo {year} {2023})}\BibitemShut {NoStop}%
\bibitem [{\citenamefont {So}(2025)}]{So+2025}%
  \BibitemOpen
  \bibfield  {author} {\bibinfo {author} {\bibfnamefont {Jae-Pil}\ \bibnamefont {So}},\ }\bibfield  {title} {\enquote {\bibinfo {title} {Deterministic generation and nanophotonic integration of 2d quantum emitters for advanced quantum photonic functionalities},}\ }\href@noop {} {\bibfield  {journal} {\bibinfo  {journal} {Nanophotonics}\ } (\bibinfo {year} {2025})}\BibitemShut {NoStop}%
\bibitem [{\citenamefont {Esmann}\ \emph {et~al.}(2024)\citenamefont {Esmann}, \citenamefont {Wein},\ and\ \citenamefont {Ant{\'o}n-Solanas}}]{esmann2024solid}%
  \BibitemOpen
  \bibfield  {author} {\bibinfo {author} {\bibfnamefont {Martin}\ \bibnamefont {Esmann}}, \bibinfo {author} {\bibfnamefont {Stephen~C.}\ \bibnamefont {Wein}}, \ and\ \bibinfo {author} {\bibfnamefont {Carlos}\ \bibnamefont {Ant{\'o}n-Solanas}},\ }\bibfield  {title} {\enquote {\bibinfo {title} {Solid-state single-photon sources: Recent advances for novel quantum materials},}\ }\href@noop {} {\bibfield  {journal} {\bibinfo  {journal} {Advanced Functional Materials}\ }\textbf {\bibinfo {volume} {34}},\ \bibinfo {pages} {2315936} (\bibinfo {year} {2024})}\BibitemShut {NoStop}%
\bibitem [{\citenamefont {Liu}\ \emph {et~al.}(2015)\citenamefont {Liu}, \citenamefont {Jiao}, \citenamefont {Xie}, \citenamefont {Yang}, \citenamefont {Chen}, \citenamefont {Ho}, \citenamefont {Gao}, \citenamefont {Jia}, \citenamefont {Cui},\ and\ \citenamefont {Xie}}]{liu2015molecular}%
  \BibitemOpen
  \bibfield  {author} {\bibinfo {author} {\bibfnamefont {H.J.}\ \bibnamefont {Liu}}, \bibinfo {author} {\bibfnamefont {L.}~\bibnamefont {Jiao}}, \bibinfo {author} {\bibfnamefont {L.}~\bibnamefont {Xie}}, \bibinfo {author} {\bibfnamefont {F.}~\bibnamefont {Yang}}, \bibinfo {author} {\bibfnamefont {J.L.}\ \bibnamefont {Chen}}, \bibinfo {author} {\bibfnamefont {W.K.}\ \bibnamefont {Ho}}, \bibinfo {author} {\bibfnamefont {C.L.}\ \bibnamefont {Gao}}, \bibinfo {author} {\bibfnamefont {J.F.}\ \bibnamefont {Jia}}, \bibinfo {author} {\bibfnamefont {X.D.}\ \bibnamefont {Cui}}, \ and\ \bibinfo {author} {\bibfnamefont {M.H.}\ \bibnamefont {Xie}},\ }\bibfield  {title} {\enquote {\bibinfo {title} {Molecular-beam epitaxy of monolayer and bilayer {WSe\(_2\)}: a scanning tunneling microscopy/spectroscopy study and deduction of exciton binding energy},}\ }\href@noop {} {\bibfield  {journal} {\bibinfo  {journal} {2D Materials}\ }\textbf {\bibinfo {volume} {2}},\ \bibinfo {pages} {034004} (\bibinfo {year} {2015})}\BibitemShut
  {NoStop}%
\bibitem [{\citenamefont {Tran}\ \emph {et~al.}(2019)\citenamefont {Tran}, \citenamefont {Moody}, \citenamefont {Wu}, \citenamefont {Lu}, \citenamefont {Choi}, \citenamefont {Kim}, \citenamefont {Rai}, \citenamefont {Sanchez}, \citenamefont {Quan}, \citenamefont {Singh} \emph {et~al.}}]{tran2019evidence}%
  \BibitemOpen
  \bibfield  {author} {\bibinfo {author} {\bibfnamefont {Kha}\ \bibnamefont {Tran}}, \bibinfo {author} {\bibfnamefont {Galan}\ \bibnamefont {Moody}}, \bibinfo {author} {\bibfnamefont {Fengcheng}\ \bibnamefont {Wu}}, \bibinfo {author} {\bibfnamefont {Xiaobo}\ \bibnamefont {Lu}}, \bibinfo {author} {\bibfnamefont {Junho}\ \bibnamefont {Choi}}, \bibinfo {author} {\bibfnamefont {Kyounghwan}\ \bibnamefont {Kim}}, \bibinfo {author} {\bibfnamefont {Amritesh}\ \bibnamefont {Rai}}, \bibinfo {author} {\bibfnamefont {Daniel~A.}\ \bibnamefont {Sanchez}}, \bibinfo {author} {\bibfnamefont {Jiamin}\ \bibnamefont {Quan}}, \bibinfo {author} {\bibfnamefont {Akshay}\ \bibnamefont {Singh}},  \emph {et~al.},\ }\bibfield  {title} {\enquote {\bibinfo {title} {Evidence for moir{\'e} excitons in van der waals heterostructures},}\ }\href@noop {} {\bibfield  {journal} {\bibinfo  {journal} {Nature}\ }\textbf {\bibinfo {volume} {567}},\ \bibinfo {pages} {71--75} (\bibinfo {year} {2019})}\BibitemShut {NoStop}%
\bibitem [{\citenamefont {Singh}\ \emph {et~al.}(2016)\citenamefont {Singh}, \citenamefont {Moody}, \citenamefont {Tran}, \citenamefont {Scott}, \citenamefont {Overbeck}, \citenamefont {Bergh{\"a}user}, \citenamefont {Schaibley}, \citenamefont {Seifert}, \citenamefont {Pleskot}, \citenamefont {Gabor} \emph {et~al.}}]{singh2016trion}%
  \BibitemOpen
  \bibfield  {author} {\bibinfo {author} {\bibfnamefont {Akshay}\ \bibnamefont {Singh}}, \bibinfo {author} {\bibfnamefont {Galan}\ \bibnamefont {Moody}}, \bibinfo {author} {\bibfnamefont {Kha}\ \bibnamefont {Tran}}, \bibinfo {author} {\bibfnamefont {Marie~E.}\ \bibnamefont {Scott}}, \bibinfo {author} {\bibfnamefont {Vincent}\ \bibnamefont {Overbeck}}, \bibinfo {author} {\bibfnamefont {Gunnar}\ \bibnamefont {Bergh{\"a}user}}, \bibinfo {author} {\bibfnamefont {John}\ \bibnamefont {Schaibley}}, \bibinfo {author} {\bibfnamefont {Edward~J.}\ \bibnamefont {Seifert}}, \bibinfo {author} {\bibfnamefont {Dennis}\ \bibnamefont {Pleskot}}, \bibinfo {author} {\bibfnamefont {Nathaniel~M}\ \bibnamefont {Gabor}},  \emph {et~al.},\ }\bibfield  {title} {\enquote {\bibinfo {title} {Trion formation dynamics in monolayer transition metal dichalcogenides},}\ }\href@noop {} {\bibfield  {journal} {\bibinfo  {journal} {Physical Review B}\ }\textbf {\bibinfo {volume} {93}},\ \bibinfo {pages} {041401} (\bibinfo {year}
  {2016})}\BibitemShut {NoStop}%
\bibitem [{\citenamefont {Tonndorf}\ \emph {et~al.}(2017)\citenamefont {Tonndorf}, \citenamefont {Del Pozo-Zamudio}, \citenamefont {Gruhler}, \citenamefont {Kern}, \citenamefont {Schmidt}, \citenamefont {Dmitriev}, \citenamefont {Bakhtinov}, \citenamefont {Tartakovskii}, \citenamefont {Pernice}, \citenamefont {Michaelis~de Vasconcellos} \emph {et~al.}}]{tonndorf2017chip}%
  \BibitemOpen
  \bibfield  {author} {\bibinfo {author} {\bibfnamefont {Philipp}\ \bibnamefont {Tonndorf}}, \bibinfo {author} {\bibfnamefont {Osvaldo}\ \bibnamefont {Del Pozo-Zamudio}}, \bibinfo {author} {\bibfnamefont {Nico}\ \bibnamefont {Gruhler}}, \bibinfo {author} {\bibfnamefont {Johannes}\ \bibnamefont {Kern}}, \bibinfo {author} {\bibfnamefont {Robert}\ \bibnamefont {Schmidt}}, \bibinfo {author} {\bibfnamefont {Alexander~I}\ \bibnamefont {Dmitriev}}, \bibinfo {author} {\bibfnamefont {Anatoly~P}\ \bibnamefont {Bakhtinov}}, \bibinfo {author} {\bibfnamefont {Alexander~I}\ \bibnamefont {Tartakovskii}}, \bibinfo {author} {\bibfnamefont {Wolfram}\ \bibnamefont {Pernice}}, \bibinfo {author} {\bibfnamefont {Steffen}\ \bibnamefont {Michaelis~de Vasconcellos}},  \emph {et~al.},\ }\bibfield  {title} {\enquote {\bibinfo {title} {On-chip waveguide coupling of a layered semiconductor single-photon source},}\ }\href@noop {} {\bibfield  {journal} {\bibinfo  {journal} {Nano Letters}\ }\textbf {\bibinfo {volume} {17}},\ \bibinfo
  {pages} {5446--5451} (\bibinfo {year} {2017})}\BibitemShut {NoStop}%
\bibitem [{\citenamefont {Aharonovich}\ \emph {et~al.}(2016)\citenamefont {Aharonovich}, \citenamefont {Englund},\ and\ \citenamefont {Toth}}]{Aharonovich2016Solid-stateEmitters}%
  \BibitemOpen
  \bibfield  {author} {\bibinfo {author} {\bibfnamefont {Igor}\ \bibnamefont {Aharonovich}}, \bibinfo {author} {\bibfnamefont {Dirk}\ \bibnamefont {Englund}}, \ and\ \bibinfo {author} {\bibfnamefont {Milos}\ \bibnamefont {Toth}},\ }\bibfield  {title} {\enquote {\bibinfo {title} {Solid-state single-photon emitters},}\ }\href {\doibase 10.1038/nphoton.2016.186} {\bibfield  {journal} {\bibinfo  {journal} {Nature Photonics}\ }\textbf {\bibinfo {volume} {10}},\ \bibinfo {pages} {631--641} (\bibinfo {year} {2016})}\BibitemShut {NoStop}%
\bibitem [{\citenamefont {Linhart}\ \emph {et~al.}(2019)\citenamefont {Linhart}, \citenamefont {Paur}, \citenamefont {Smejkal}, \citenamefont {Burgd\"orfer}, \citenamefont {Mueller},\ and\ \citenamefont {Libisch}}]{linhart}%
  \BibitemOpen
  \bibfield  {author} {\bibinfo {author} {\bibfnamefont {Lukas}\ \bibnamefont {Linhart}}, \bibinfo {author} {\bibfnamefont {Matthias}\ \bibnamefont {Paur}}, \bibinfo {author} {\bibfnamefont {Valerie}\ \bibnamefont {Smejkal}}, \bibinfo {author} {\bibfnamefont {Joachim}\ \bibnamefont {Burgd\"orfer}}, \bibinfo {author} {\bibfnamefont {Thomas}\ \bibnamefont {Mueller}}, \ and\ \bibinfo {author} {\bibfnamefont {Florian}\ \bibnamefont {Libisch}},\ }\bibfield  {title} {\enquote {\bibinfo {title} {Localized intervalley defect excitons as single-photon emitters in {WSe\(_2\)}},}\ }\href {\doibase 10.1103/PhysRevLett.123.146401} {\bibfield  {journal} {\bibinfo  {journal} {Physical Review Letter}\ }\textbf {\bibinfo {volume} {123}},\ \bibinfo {pages} {146401} (\bibinfo {year} {2019})}\BibitemShut {NoStop}%
\bibitem [{\citenamefont {Stevens}\ \emph {et~al.}(2022)\citenamefont {Stevens}, \citenamefont {Chuang}, \citenamefont {Rosenberger}, \citenamefont {McCreary}, \citenamefont {Dass}, \citenamefont {Jonker},\ and\ \citenamefont {Hendrickson}}]{stevens2022enhancing}%
  \BibitemOpen
  \bibfield  {author} {\bibinfo {author} {\bibfnamefont {Christopher~E.}\ \bibnamefont {Stevens}}, \bibinfo {author} {\bibfnamefont {Hsun-Jen}\ \bibnamefont {Chuang}}, \bibinfo {author} {\bibfnamefont {Matthew~R.}\ \bibnamefont {Rosenberger}}, \bibinfo {author} {\bibfnamefont {Kathleen~M.}\ \bibnamefont {McCreary}}, \bibinfo {author} {\bibfnamefont {Chandriker~Kavir}\ \bibnamefont {Dass}}, \bibinfo {author} {\bibfnamefont {Berend~T.}\ \bibnamefont {Jonker}}, \ and\ \bibinfo {author} {\bibfnamefont {Joshua~R.}\ \bibnamefont {Hendrickson}},\ }\bibfield  {title} {\enquote {\bibinfo {title} {Enhancing the purity of deterministically placed quantum emitters in monolayer {WSe\(_2\)}},}\ }\href@noop {} {\bibfield  {journal} {\bibinfo  {journal} {ACS Nano}\ }\textbf {\bibinfo {volume} {16}},\ \bibinfo {pages} {20956--20963} (\bibinfo {year} {2022})}\BibitemShut {NoStop}%
\bibitem [{\citenamefont {Abramov}\ \emph {et~al.}(2023)\citenamefont {Abramov}, \citenamefont {Chestnov}, \citenamefont {Alimova}, \citenamefont {Ivanova}, \citenamefont {Mukhin}, \citenamefont {Krizhanovskii}, \citenamefont {Shelykh}, \citenamefont {Iorsh},\ and\ \citenamefont {Kravtsov}}]{abramov2023photoluminescence}%
  \BibitemOpen
  \bibfield  {author} {\bibinfo {author} {\bibfnamefont {Artem~N.}\ \bibnamefont {Abramov}}, \bibinfo {author} {\bibfnamefont {Igor~Y.}\ \bibnamefont {Chestnov}}, \bibinfo {author} {\bibfnamefont {Ekaterina~S.}\ \bibnamefont {Alimova}}, \bibinfo {author} {\bibfnamefont {Tatiana}\ \bibnamefont {Ivanova}}, \bibinfo {author} {\bibfnamefont {Ivan~S.}\ \bibnamefont {Mukhin}}, \bibinfo {author} {\bibfnamefont {Dmitry~N.}\ \bibnamefont {Krizhanovskii}}, \bibinfo {author} {\bibfnamefont {Ivan~A.}\ \bibnamefont {Shelykh}}, \bibinfo {author} {\bibfnamefont {Ivan~V.}\ \bibnamefont {Iorsh}}, \ and\ \bibinfo {author} {\bibfnamefont {Vasily}\ \bibnamefont {Kravtsov}},\ }\bibfield  {title} {\enquote {\bibinfo {title} {Photoluminescence imaging of single photon emitters within nanoscale strain profiles in monolayer {WSe\(_2\)}},}\ }\href@noop {} {\bibfield  {journal} {\bibinfo  {journal} {Nature Communications}\ }\textbf {\bibinfo {volume} {14}},\ \bibinfo {pages} {5737} (\bibinfo {year} {2023})}\BibitemShut {NoStop}%
\bibitem [{\citenamefont {Paralikis}\ \emph {et~al.}(2024)\citenamefont {Paralikis}, \citenamefont {Piccinini}, \citenamefont {Madigawa}, \citenamefont {Metuh}, \citenamefont {Vannucci}, \citenamefont {Gregersen},\ and\ \citenamefont {Munkhbat}}]{paralikis2024tailoring}%
  \BibitemOpen
  \bibfield  {author} {\bibinfo {author} {\bibfnamefont {Athanasios}\ \bibnamefont {Paralikis}}, \bibinfo {author} {\bibfnamefont {Claudia}\ \bibnamefont {Piccinini}}, \bibinfo {author} {\bibfnamefont {Abdulmalik~A.}\ \bibnamefont {Madigawa}}, \bibinfo {author} {\bibfnamefont {Pietro}\ \bibnamefont {Metuh}}, \bibinfo {author} {\bibfnamefont {Luca}\ \bibnamefont {Vannucci}}, \bibinfo {author} {\bibfnamefont {Niels}\ \bibnamefont {Gregersen}}, \ and\ \bibinfo {author} {\bibfnamefont {Battulga}\ \bibnamefont {Munkhbat}},\ }\bibfield  {title} {\enquote {\bibinfo {title} {Tailoring polarization in {WSe\(_2\)} quantum emitters through deterministic strain engineering},}\ }\href@noop {} {\bibfield  {journal} {\bibinfo  {journal} {npj 2D Materials and Applications}\ }\textbf {\bibinfo {volume} {8}},\ \bibinfo {pages} {59} (\bibinfo {year} {2024})}\BibitemShut {NoStop}%
\bibitem [{\citenamefont {Parto}\ \emph {et~al.}(2021)\citenamefont {Parto}, \citenamefont {Azzam}, \citenamefont {Banerjee},\ and\ \citenamefont {Moody}}]{Parto2021DefectK}%
  \BibitemOpen
  \bibfield  {author} {\bibinfo {author} {\bibfnamefont {Kamyar}\ \bibnamefont {Parto}}, \bibinfo {author} {\bibfnamefont {Shaimaa~I.}\ \bibnamefont {Azzam}}, \bibinfo {author} {\bibfnamefont {Kaustav}\ \bibnamefont {Banerjee}}, \ and\ \bibinfo {author} {\bibfnamefont {Galan}\ \bibnamefont {Moody}},\ }\bibfield  {title} {\enquote {\bibinfo {title} {Defect and strain engineering of monolayer {WSe\(_2\)} enables site-controlled single-photon emission up to 150 k},}\ }\href {\doibase 10.1038/s41467-021-23709-5} {\bibfield  {journal} {\bibinfo  {journal} {Nature Communications}\ }\textbf {\bibinfo {volume} {12}} (\bibinfo {year} {2021}),\ 10.1038/s41467-021-23709-5}\BibitemShut {NoStop}%
\bibitem [{\citenamefont {Peng}\ \emph {et~al.}(2020)\citenamefont {Peng}, \citenamefont {Chan}, \citenamefont {Choo}, \citenamefont {Odom}, \citenamefont {Sankaranarayanan},\ and\ \citenamefont {Ma}}]{peng2020creation}%
  \BibitemOpen
  \bibfield  {author} {\bibinfo {author} {\bibfnamefont {Lintao}\ \bibnamefont {Peng}}, \bibinfo {author} {\bibfnamefont {Henry}\ \bibnamefont {Chan}}, \bibinfo {author} {\bibfnamefont {Priscilla}\ \bibnamefont {Choo}}, \bibinfo {author} {\bibfnamefont {Teri~W.}\ \bibnamefont {Odom}}, \bibinfo {author} {\bibfnamefont {Subramanian~K.R.S.}\ \bibnamefont {Sankaranarayanan}}, \ and\ \bibinfo {author} {\bibfnamefont {Xuedan}\ \bibnamefont {Ma}},\ }\bibfield  {title} {\enquote {\bibinfo {title} {Creation of single-photon emitters in {WSe\(_2\)} monolayers using nanometer-sized gold tips},}\ }\href@noop {} {\bibfield  {journal} {\bibinfo  {journal} {Nano Letters}\ }\textbf {\bibinfo {volume} {20}},\ \bibinfo {pages} {5866--5872} (\bibinfo {year} {2020})}\BibitemShut {NoStop}%
\bibitem [{\citenamefont {Azzam}\ \emph {et~al.}(2023)\citenamefont {Azzam}, \citenamefont {Parto},\ and\ \citenamefont {Moody}}]{azzam2023purcell}%
  \BibitemOpen
  \bibfield  {author} {\bibinfo {author} {\bibfnamefont {Shaimaa~I.}\ \bibnamefont {Azzam}}, \bibinfo {author} {\bibfnamefont {Kamyar}\ \bibnamefont {Parto}}, \ and\ \bibinfo {author} {\bibfnamefont {Galan}\ \bibnamefont {Moody}},\ }\bibfield  {title} {\enquote {\bibinfo {title} {Purcell enhancement and polarization control of single-photon emitters in monolayer {WSe\(_2\)} using dielectric nanoantennas},}\ }\href@noop {} {\bibfield  {journal} {\bibinfo  {journal} {Nanophotonics}\ }\textbf {\bibinfo {volume} {12}},\ \bibinfo {pages} {477--484} (\bibinfo {year} {2023})}\BibitemShut {NoStop}%
\bibitem [{\citenamefont {Luo}\ \emph {et~al.}(2018)\citenamefont {Luo}, \citenamefont {Shepard}, \citenamefont {Ardelean}, \citenamefont {Rhodes}, \citenamefont {Kim}, \citenamefont {Barmak}, \citenamefont {Hone},\ and\ \citenamefont {Strauf}}]{luo2018deterministic}%
  \BibitemOpen
  \bibfield  {author} {\bibinfo {author} {\bibfnamefont {Yue}\ \bibnamefont {Luo}}, \bibinfo {author} {\bibfnamefont {Gabriella~D.}\ \bibnamefont {Shepard}}, \bibinfo {author} {\bibfnamefont {Jenny~V.}\ \bibnamefont {Ardelean}}, \bibinfo {author} {\bibfnamefont {Daniel~A.}\ \bibnamefont {Rhodes}}, \bibinfo {author} {\bibfnamefont {Bumho}\ \bibnamefont {Kim}}, \bibinfo {author} {\bibfnamefont {Katayun}\ \bibnamefont {Barmak}}, \bibinfo {author} {\bibfnamefont {James~C.}\ \bibnamefont {Hone}}, \ and\ \bibinfo {author} {\bibfnamefont {Stefan}\ \bibnamefont {Strauf}},\ }\bibfield  {title} {\enquote {\bibinfo {title} {Deterministic coupling of site-controlled quantum emitters in monolayer {WSe\(_2\)} to plasmonic nanocavities},}\ }\href@noop {} {\bibfield  {journal} {\bibinfo  {journal} {Nature Nanotechnology}\ }\textbf {\bibinfo {volume} {13}},\ \bibinfo {pages} {1137--1142} (\bibinfo {year} {2018})}\BibitemShut {NoStop}%
\bibitem [{\citenamefont {Xu}\ \emph {et~al.}(2024)\citenamefont {Xu}, \citenamefont {Vong}, \citenamefont {Utama}, \citenamefont {Lebedev}, \citenamefont {Ananth}, \citenamefont {Hersam}, \citenamefont {Weiss},\ and\ \citenamefont {Mirkin}}]{xu2024sub}%
  \BibitemOpen
  \bibfield  {author} {\bibinfo {author} {\bibfnamefont {David~D}\ \bibnamefont {Xu}}, \bibinfo {author} {\bibfnamefont {Albert~F}\ \bibnamefont {Vong}}, \bibinfo {author} {\bibfnamefont {M~Iqbal~Bakti}\ \bibnamefont {Utama}}, \bibinfo {author} {\bibfnamefont {Dmitry}\ \bibnamefont {Lebedev}}, \bibinfo {author} {\bibfnamefont {Riddhi}\ \bibnamefont {Ananth}}, \bibinfo {author} {\bibfnamefont {Mark~C}\ \bibnamefont {Hersam}}, \bibinfo {author} {\bibfnamefont {Emily~A}\ \bibnamefont {Weiss}}, \ and\ \bibinfo {author} {\bibfnamefont {Chad~A}\ \bibnamefont {Mirkin}},\ }\bibfield  {title} {\enquote {\bibinfo {title} {Sub-diffraction correlation of quantum emitters and local strain fields in strain-engineered {WSe\(_2\)} monolayers},}\ }\href@noop {} {\bibfield  {journal} {\bibinfo  {journal} {Advanced Materials}\ }\textbf {\bibinfo {volume} {36}},\ \bibinfo {pages} {2314242} (\bibinfo {year} {2024})}\BibitemShut {NoStop}%
\bibitem [{\citenamefont {Xu}\ \emph {et~al.}(2023)\citenamefont {Xu}, \citenamefont {Vong}, \citenamefont {Lebedev}, \citenamefont {Ananth}, \citenamefont {Wong}, \citenamefont {Brown}, \citenamefont {Hersam}, \citenamefont {Mirkin},\ and\ \citenamefont {Weiss}}]{xu2023conversion}%
  \BibitemOpen
  \bibfield  {author} {\bibinfo {author} {\bibfnamefont {David~D.}\ \bibnamefont {Xu}}, \bibinfo {author} {\bibfnamefont {Albert~F.}\ \bibnamefont {Vong}}, \bibinfo {author} {\bibfnamefont {Dmitry}\ \bibnamefont {Lebedev}}, \bibinfo {author} {\bibfnamefont {Riddhi}\ \bibnamefont {Ananth}}, \bibinfo {author} {\bibfnamefont {Alexa~M.}\ \bibnamefont {Wong}}, \bibinfo {author} {\bibfnamefont {Paul~T.}\ \bibnamefont {Brown}}, \bibinfo {author} {\bibfnamefont {Mark~C.}\ \bibnamefont {Hersam}}, \bibinfo {author} {\bibfnamefont {Chad~A.}\ \bibnamefont {Mirkin}}, \ and\ \bibinfo {author} {\bibfnamefont {Emily~A.}\ \bibnamefont {Weiss}},\ }\bibfield  {title} {\enquote {\bibinfo {title} {Conversion of classical light emission from a nanoparticle-strained {WSe\(_2\)} monolayer into quantum light emission via electron beam irradiation},}\ }\href@noop {} {\bibfield  {journal} {\bibinfo  {journal} {Advanced Materials}\ }\textbf {\bibinfo {volume} {35}},\ \bibinfo {pages} {2208066} (\bibinfo {year} {2023})}\BibitemShut
  {NoStop}%
\bibitem [{\citenamefont {Tripathi}\ \emph {et~al.}(2018)\citenamefont {Tripathi}, \citenamefont {Iff}, \citenamefont {Betzold}, \citenamefont {Dusanowski}, \citenamefont {Emmerling}, \citenamefont {Moon}, \citenamefont {Lee}, \citenamefont {Kwon}, \citenamefont {H{\"o}fling},\ and\ \citenamefont {Schneider}}]{tripathi2018spontaneous}%
  \BibitemOpen
  \bibfield  {author} {\bibinfo {author} {\bibfnamefont {Laxmi~Narayan}\ \bibnamefont {Tripathi}}, \bibinfo {author} {\bibfnamefont {Oliver}\ \bibnamefont {Iff}}, \bibinfo {author} {\bibfnamefont {Simon}\ \bibnamefont {Betzold}}, \bibinfo {author} {\bibfnamefont {{\L}ukasz}\ \bibnamefont {Dusanowski}}, \bibinfo {author} {\bibfnamefont {Monika}\ \bibnamefont {Emmerling}}, \bibinfo {author} {\bibfnamefont {Kihwan}\ \bibnamefont {Moon}}, \bibinfo {author} {\bibfnamefont {Young~Jin}\ \bibnamefont {Lee}}, \bibinfo {author} {\bibfnamefont {Soon-Hong}\ \bibnamefont {Kwon}}, \bibinfo {author} {\bibfnamefont {Sven}\ \bibnamefont {H{\"o}fling}}, \ and\ \bibinfo {author} {\bibfnamefont {Christian}\ \bibnamefont {Schneider}},\ }\bibfield  {title} {\enquote {\bibinfo {title} {Spontaneous emission enhancement in strain-induced {WSe\(_2\)} monolayer-based quantum light sources on metallic surfaces},}\ }\href@noop {} {\bibfield  {journal} {\bibinfo  {journal} {ACS Photonics}\ }\textbf {\bibinfo {volume} {5}},\ \bibinfo
  {pages} {1919--1926} (\bibinfo {year} {2018})}\BibitemShut {NoStop}%
\bibitem [{\citenamefont {Iff}\ \emph {et~al.}(2019)\citenamefont {Iff}, \citenamefont {Tedeschi}, \citenamefont {Mart{\'\i}n-S{\'a}nchez}, \citenamefont {Mocza{\l}a-Dusanowska}, \citenamefont {Tongay}, \citenamefont {Yumigeta}, \citenamefont {Taboada-Guti{\'e}rrez}, \citenamefont {Savaresi}, \citenamefont {Rastelli}, \citenamefont {Alonso-Gonz{\'a}lez} \emph {et~al.}}]{iff2019strain}%
  \BibitemOpen
  \bibfield  {author} {\bibinfo {author} {\bibfnamefont {Oliver}\ \bibnamefont {Iff}}, \bibinfo {author} {\bibfnamefont {Davide}\ \bibnamefont {Tedeschi}}, \bibinfo {author} {\bibfnamefont {Javier}\ \bibnamefont {Mart{\'\i}n-S{\'a}nchez}}, \bibinfo {author} {\bibfnamefont {Magdalena}\ \bibnamefont {Mocza{\l}a-Dusanowska}}, \bibinfo {author} {\bibfnamefont {Sefaattin}\ \bibnamefont {Tongay}}, \bibinfo {author} {\bibfnamefont {Kentaro}\ \bibnamefont {Yumigeta}}, \bibinfo {author} {\bibfnamefont {Javier}\ \bibnamefont {Taboada-Guti{\'e}rrez}}, \bibinfo {author} {\bibfnamefont {Matteo}\ \bibnamefont {Savaresi}}, \bibinfo {author} {\bibfnamefont {Armando}\ \bibnamefont {Rastelli}}, \bibinfo {author} {\bibfnamefont {Pablo}\ \bibnamefont {Alonso-Gonz{\'a}lez}},  \emph {et~al.},\ }\bibfield  {title} {\enquote {\bibinfo {title} {Strain-tunable single photon sources in {WSe\(_2\)} monolayers},}\ }\href@noop {} {\bibfield  {journal} {\bibinfo  {journal} {Nano Letters}\ }\textbf {\bibinfo {volume} {19}},\ \bibinfo
  {pages} {6931--6936} (\bibinfo {year} {2019})}\BibitemShut {NoStop}%
\bibitem [{\citenamefont {Klein}\ \emph {et~al.}(2019)\citenamefont {Klein}, \citenamefont {Lorke}, \citenamefont {Florian}, \citenamefont {Sigger}, \citenamefont {Sigl}, \citenamefont {Rey}, \citenamefont {Wierzbowski}, \citenamefont {Cerne}, \citenamefont {M{\"u}ller}, \citenamefont {Mitterreiter} \emph {et~al.}}]{klein2019site}%
  \BibitemOpen
  \bibfield  {author} {\bibinfo {author} {\bibfnamefont {J.}~\bibnamefont {Klein}}, \bibinfo {author} {\bibfnamefont {M.}~\bibnamefont {Lorke}}, \bibinfo {author} {\bibfnamefont {M.}~\bibnamefont {Florian}}, \bibinfo {author} {\bibfnamefont {F.}~\bibnamefont {Sigger}}, \bibinfo {author} {\bibfnamefont {L.}~\bibnamefont {Sigl}}, \bibinfo {author} {\bibfnamefont {S.}~\bibnamefont {Rey}}, \bibinfo {author} {\bibfnamefont {J.}~\bibnamefont {Wierzbowski}}, \bibinfo {author} {\bibfnamefont {J.}~\bibnamefont {Cerne}}, \bibinfo {author} {\bibfnamefont {K.}~\bibnamefont {M{\"u}ller}}, \bibinfo {author} {\bibfnamefont {E.}~\bibnamefont {Mitterreiter}},  \emph {et~al.},\ }\bibfield  {title} {\enquote {\bibinfo {title} {Site-selectively generated photon emitters in monolayer {MoS\(_2\)} via local helium ion irradiation},}\ }\href@noop {} {\bibfield  {journal} {\bibinfo  {journal} {Nature Communications}\ }\textbf {\bibinfo {volume} {10}},\ \bibinfo {pages} {2755} (\bibinfo {year} {2019})}\BibitemShut {NoStop}%
\bibitem [{\citenamefont {Micevic}\ \emph {et~al.}(2022)\citenamefont {Micevic}, \citenamefont {Pettinger}, \citenamefont {H{\"o}tger}, \citenamefont {Sigl}, \citenamefont {Florian}, \citenamefont {Taniguchi}, \citenamefont {Watanabe}, \citenamefont {M{\"u}ller}, \citenamefont {Finley}, \citenamefont {Kastl} \emph {et~al.}}]{micevic2022demand}%
  \BibitemOpen
  \bibfield  {author} {\bibinfo {author} {\bibfnamefont {A.}~\bibnamefont {Micevic}}, \bibinfo {author} {\bibfnamefont {N.}~\bibnamefont {Pettinger}}, \bibinfo {author} {\bibfnamefont {A.}~\bibnamefont {H{\"o}tger}}, \bibinfo {author} {\bibfnamefont {L.}~\bibnamefont {Sigl}}, \bibinfo {author} {\bibfnamefont {M.}~\bibnamefont {Florian}}, \bibinfo {author} {\bibfnamefont {T.}~\bibnamefont {Taniguchi}}, \bibinfo {author} {\bibfnamefont {K.}~\bibnamefont {Watanabe}}, \bibinfo {author} {\bibfnamefont {K.}~\bibnamefont {M{\"u}ller}}, \bibinfo {author} {\bibfnamefont {J.J.}\ \bibnamefont {Finley}}, \bibinfo {author} {\bibfnamefont {C.}~\bibnamefont {Kastl}},  \emph {et~al.},\ }\bibfield  {title} {\enquote {\bibinfo {title} {On-demand generation of optically active defects in monolayer {WS\(_2\)} by a focused helium ion beam},}\ }\href@noop {} {\bibfield  {journal} {\bibinfo  {journal} {Applied Physics Letters}\ }\textbf {\bibinfo {volume} {121}} (\bibinfo {year} {2022})}\BibitemShut {NoStop}%
\bibitem [{\citenamefont {Wang}\ \emph {et~al.}(2022)\citenamefont {Wang}, \citenamefont {Jones}, \citenamefont {Chen}, \citenamefont {Schatz},\ and\ \citenamefont {Ma}}]{wang2022utilizing}%
  \BibitemOpen
  \bibfield  {author} {\bibinfo {author} {\bibfnamefont {Wei}\ \bibnamefont {Wang}}, \bibinfo {author} {\bibfnamefont {Leighton~O.}\ \bibnamefont {Jones}}, \bibinfo {author} {\bibfnamefont {Jia-Shiang}\ \bibnamefont {Chen}}, \bibinfo {author} {\bibfnamefont {George~C.}\ \bibnamefont {Schatz}}, \ and\ \bibinfo {author} {\bibfnamefont {Xuedan}\ \bibnamefont {Ma}},\ }\bibfield  {title} {\enquote {\bibinfo {title} {Utilizing ultraviolet photons to generate single-photon emitters in semiconductor monolayers},}\ }\href@noop {} {\bibfield  {journal} {\bibinfo  {journal} {ACS Nano}\ }\textbf {\bibinfo {volume} {16}},\ \bibinfo {pages} {21240--21247} (\bibinfo {year} {2022})}\BibitemShut {NoStop}%
\bibitem [{\citenamefont {Zhao}\ \emph {et~al.}(2021)\citenamefont {Zhao}, \citenamefont {Pettes}, \citenamefont {Zheng},\ and\ \citenamefont {Htoon}}]{zhao2021site}%
  \BibitemOpen
  \bibfield  {author} {\bibinfo {author} {\bibfnamefont {Huan}\ \bibnamefont {Zhao}}, \bibinfo {author} {\bibfnamefont {Michael~T}\ \bibnamefont {Pettes}}, \bibinfo {author} {\bibfnamefont {Yu}~\bibnamefont {Zheng}}, \ and\ \bibinfo {author} {\bibfnamefont {Han}\ \bibnamefont {Htoon}},\ }\bibfield  {title} {\enquote {\bibinfo {title} {Site-controlled telecom-wavelength single-photon emitters in atomically-thin {MoTe\(_2\)}},}\ }\href@noop {} {\bibfield  {journal} {\bibinfo  {journal} {Nature Communications}\ }\textbf {\bibinfo {volume} {12}},\ \bibinfo {pages} {6753} (\bibinfo {year} {2021})}\BibitemShut {NoStop}%
\bibitem [{\citenamefont {Yu}\ \emph {et~al.}(2021)\citenamefont {Yu}, \citenamefont {Deng}, \citenamefont {Zhang}, \citenamefont {Borghardt}, \citenamefont {Kardynal}, \citenamefont {Vuckovic},\ and\ \citenamefont {Heinz}}]{yu2021site}%
  \BibitemOpen
  \bibfield  {author} {\bibinfo {author} {\bibfnamefont {Leo}\ \bibnamefont {Yu}}, \bibinfo {author} {\bibfnamefont {Minda}\ \bibnamefont {Deng}}, \bibinfo {author} {\bibfnamefont {Jingyuan~Linda}\ \bibnamefont {Zhang}}, \bibinfo {author} {\bibfnamefont {Sven}\ \bibnamefont {Borghardt}}, \bibinfo {author} {\bibfnamefont {Beata}\ \bibnamefont {Kardynal}}, \bibinfo {author} {\bibfnamefont {Jelena}\ \bibnamefont {Vuckovic}}, \ and\ \bibinfo {author} {\bibfnamefont {Tony~F}\ \bibnamefont {Heinz}},\ }\bibfield  {title} {\enquote {\bibinfo {title} {Site-controlled quantum emitters in monolayer {MoSe\(_2\)}},}\ }\href@noop {} {\bibfield  {journal} {\bibinfo  {journal} {Nano Letters}\ }\textbf {\bibinfo {volume} {21}},\ \bibinfo {pages} {2376--2381} (\bibinfo {year} {2021})}\BibitemShut {NoStop}%
\bibitem [{\citenamefont {Lenferink}\ \emph {et~al.}(2022)\citenamefont {Lenferink}, \citenamefont {LaMountain}, \citenamefont {Stanev}, \citenamefont {Garvey}, \citenamefont {Watanabe}, \citenamefont {Taniguchi},\ and\ \citenamefont {Stern}}]{lenferink2022tunable}%
  \BibitemOpen
  \bibfield  {author} {\bibinfo {author} {\bibfnamefont {Erik~J.}\ \bibnamefont {Lenferink}}, \bibinfo {author} {\bibfnamefont {Trevor}\ \bibnamefont {LaMountain}}, \bibinfo {author} {\bibfnamefont {Teodor~K.}\ \bibnamefont {Stanev}}, \bibinfo {author} {\bibfnamefont {Ethan}\ \bibnamefont {Garvey}}, \bibinfo {author} {\bibfnamefont {Kenji}\ \bibnamefont {Watanabe}}, \bibinfo {author} {\bibfnamefont {Takashi}\ \bibnamefont {Taniguchi}}, \ and\ \bibinfo {author} {\bibfnamefont {Nathaniel~P.}\ \bibnamefont {Stern}},\ }\bibfield  {title} {\enquote {\bibinfo {title} {Tunable emission from localized excitons deterministically positioned in monolayer p-n junctions},}\ }\href@noop {} {\bibfield  {journal} {\bibinfo  {journal} {ACS Photonics}\ }\textbf {\bibinfo {volume} {9}},\ \bibinfo {pages} {3067--3074} (\bibinfo {year} {2022})}\BibitemShut {NoStop}%
\bibitem [{\citenamefont {Chakraborty}\ \emph {et~al.}(2019)\citenamefont {Chakraborty}, \citenamefont {Jungwirth}, \citenamefont {Fuchs},\ and\ \citenamefont {Vamivakas}}]{chakraborty2019electrical}%
  \BibitemOpen
  \bibfield  {author} {\bibinfo {author} {\bibfnamefont {Chitraleema}\ \bibnamefont {Chakraborty}}, \bibinfo {author} {\bibfnamefont {Nicholas~R.}\ \bibnamefont {Jungwirth}}, \bibinfo {author} {\bibfnamefont {Gregory~D.}\ \bibnamefont {Fuchs}}, \ and\ \bibinfo {author} {\bibfnamefont {A.~Nick}\ \bibnamefont {Vamivakas}},\ }\bibfield  {title} {\enquote {\bibinfo {title} {Electrical manipulation of the fine-structure splitting of {WSe\(_2\)} quantum emitters},}\ }\href@noop {} {\bibfield  {journal} {\bibinfo  {journal} {Physical Review B}\ }\textbf {\bibinfo {volume} {99}},\ \bibinfo {pages} {045308} (\bibinfo {year} {2019})}\BibitemShut {NoStop}%
\bibitem [{\citenamefont {Howarth}\ \emph {et~al.}(2024)\citenamefont {Howarth}, \citenamefont {Vaklinova}, \citenamefont {Grzeszczyk}, \citenamefont {Baldi}, \citenamefont {Hague}, \citenamefont {Potemski}, \citenamefont {Novoselov}, \citenamefont {Kozikov},\ and\ \citenamefont {Koperski}}]{howarth2024electroluminescent}%
  \BibitemOpen
  \bibfield  {author} {\bibinfo {author} {\bibfnamefont {James}\ \bibnamefont {Howarth}}, \bibinfo {author} {\bibfnamefont {Kristina}\ \bibnamefont {Vaklinova}}, \bibinfo {author} {\bibfnamefont {Magdalena}\ \bibnamefont {Grzeszczyk}}, \bibinfo {author} {\bibfnamefont {Giulio}\ \bibnamefont {Baldi}}, \bibinfo {author} {\bibfnamefont {Lee}\ \bibnamefont {Hague}}, \bibinfo {author} {\bibfnamefont {Marek}\ \bibnamefont {Potemski}}, \bibinfo {author} {\bibfnamefont {Kostya~S.}\ \bibnamefont {Novoselov}}, \bibinfo {author} {\bibfnamefont {Aleksey}\ \bibnamefont {Kozikov}}, \ and\ \bibinfo {author} {\bibfnamefont {Maciej}\ \bibnamefont {Koperski}},\ }\bibfield  {title} {\enquote {\bibinfo {title} {Electroluminescent vertical tunneling junctions based on {WSe\(_2\)} monolayer quantum emitter arrays: Exploring tunability with electric and magnetic fields},}\ }\href@noop {} {\bibfield  {journal} {\bibinfo  {journal} {Proceedings of the National Academy of Sciences}\ }\textbf {\bibinfo {volume} {121}},\ \bibinfo
  {pages} {e2401757121} (\bibinfo {year} {2024})}\BibitemShut {NoStop}%
\bibitem [{\citenamefont {Drawer}\ \emph {et~al.}(2023)\citenamefont {Drawer}, \citenamefont {Mitryakhin}, \citenamefont {Shan}, \citenamefont {Stephan}, \citenamefont {Gittinger}, \citenamefont {Lackner}, \citenamefont {Han}, \citenamefont {Leibeling}, \citenamefont {Eilenberger},\ and\ \citenamefont {Banerjee}}]{drawer2023monolayer}%
  \BibitemOpen
  \bibfield  {author} {\bibinfo {author} {\bibfnamefont {Jens-Christian}\ \bibnamefont {Drawer}}, \bibinfo {author} {\bibfnamefont {Victor~Nikolaevich}\ \bibnamefont {Mitryakhin}}, \bibinfo {author} {\bibfnamefont {Hangyong}\ \bibnamefont {Shan}}, \bibinfo {author} {\bibfnamefont {Sven}\ \bibnamefont {Stephan}}, \bibinfo {author} {\bibfnamefont {Moritz}\ \bibnamefont {Gittinger}}, \bibinfo {author} {\bibfnamefont {Lukas}\ \bibnamefont {Lackner}}, \bibinfo {author} {\bibfnamefont {Bo}~\bibnamefont {Han}}, \bibinfo {author} {\bibfnamefont {Gilbert}\ \bibnamefont {Leibeling}}, \bibinfo {author} {\bibfnamefont {Falk}\ \bibnamefont {Eilenberger}}, \ and\ \bibinfo {author} {\bibfnamefont {Rounak et~al.}\ \bibnamefont {Banerjee}},\ }\bibfield  {title} {\enquote {\bibinfo {title} {Monolayer-based single-photon source in a liquid-helium-free open cavity featuring 65\% brightness and quantum coherence},}\ }\href@noop {} {\bibfield  {journal} {\bibinfo  {journal} {Nano Letters}\ }\textbf {\bibinfo {volume} {23}},\
  \bibinfo {pages} {8683--8689} (\bibinfo {year} {2023})}\BibitemShut {NoStop}%
\bibitem [{\citenamefont {Vannucci}\ \emph {et~al.}(2024)\citenamefont {Vannucci}, \citenamefont {Neto}, \citenamefont {Piccinini}, \citenamefont {Paralikis}, \citenamefont {Gregersen},\ and\ \citenamefont {Munkhbat}}]{vannucci2024single}%
  \BibitemOpen
  \bibfield  {author} {\bibinfo {author} {\bibfnamefont {Luca}\ \bibnamefont {Vannucci}}, \bibinfo {author} {\bibfnamefont {Jos{\'e}~Ferreira}\ \bibnamefont {Neto}}, \bibinfo {author} {\bibfnamefont {Claudia}\ \bibnamefont {Piccinini}}, \bibinfo {author} {\bibfnamefont {Athanasios}\ \bibnamefont {Paralikis}}, \bibinfo {author} {\bibfnamefont {Niels}\ \bibnamefont {Gregersen}}, \ and\ \bibinfo {author} {\bibfnamefont {Battulga}\ \bibnamefont {Munkhbat}},\ }\bibfield  {title} {\enquote {\bibinfo {title} {Single-photon emitters in {WSe\(_2\)}: Critical role of phonons on excitation schemes and indistinguishability},}\ }\href@noop {} {\bibfield  {journal} {\bibinfo  {journal} {Physical Review B}\ }\textbf {\bibinfo {volume} {109}},\ \bibinfo {pages} {245304} (\bibinfo {year} {2024})}\BibitemShut {NoStop}%
\bibitem [{\citenamefont {Srivastava}\ \emph {et~al.}(2015)\citenamefont {Srivastava}, \citenamefont {Sidler}, \citenamefont {Allain}, \citenamefont {Lembke}, \citenamefont {Kis},\ and\ \citenamefont {{\.I}mamo{\u{g}}lu}}]{srivastava2015optically}%
  \BibitemOpen
  \bibfield  {author} {\bibinfo {author} {\bibfnamefont {Ajit}\ \bibnamefont {Srivastava}}, \bibinfo {author} {\bibfnamefont {Meinrad}\ \bibnamefont {Sidler}}, \bibinfo {author} {\bibfnamefont {Adrien~V.}\ \bibnamefont {Allain}}, \bibinfo {author} {\bibfnamefont {Dominik~S.}\ \bibnamefont {Lembke}}, \bibinfo {author} {\bibfnamefont {Andras}\ \bibnamefont {Kis}}, \ and\ \bibinfo {author} {\bibfnamefont {Ata{\c{c}}}\ \bibnamefont {{\.I}mamo{\u{g}}lu}},\ }\bibfield  {title} {\enquote {\bibinfo {title} {Optically active quantum dots in monolayer {WSe\(_2\)}},}\ }\href@noop {} {\bibfield  {journal} {\bibinfo  {journal} {Nature Nanotechnology}\ }\textbf {\bibinfo {volume} {10}},\ \bibinfo {pages} {491--496} (\bibinfo {year} {2015})}\BibitemShut {NoStop}%
\bibitem [{\citenamefont {Kumar}\ \emph {et~al.}(2015)\citenamefont {Kumar}, \citenamefont {Kaczmarczyk},\ and\ \citenamefont {Gerardot}}]{kumar2015strain}%
  \BibitemOpen
  \bibfield  {author} {\bibinfo {author} {\bibfnamefont {Santosh}\ \bibnamefont {Kumar}}, \bibinfo {author} {\bibfnamefont {Artur}\ \bibnamefont {Kaczmarczyk}}, \ and\ \bibinfo {author} {\bibfnamefont {Brian~D.}\ \bibnamefont {Gerardot}},\ }\bibfield  {title} {\enquote {\bibinfo {title} {Strain-induced spatial and spectral isolation of quantum emitters in mono-and bilayer {WSe\(_2\)}},}\ }\href@noop {} {\bibfield  {journal} {\bibinfo  {journal} {Nano Letters}\ }\textbf {\bibinfo {volume} {15}},\ \bibinfo {pages} {7567--7573} (\bibinfo {year} {2015})}\BibitemShut {NoStop}%
\bibitem [{\citenamefont {Piccinini}\ \emph {et~al.}(2025)\citenamefont {Piccinini}, \citenamefont {Paralikis}, \citenamefont {Neto}, \citenamefont {Madigawa}, \citenamefont {Wyborski}, \citenamefont {Remesh}, \citenamefont {Vannucci}, \citenamefont {Gregersen},\ and\ \citenamefont {Munkhbat}}]{piccinini2024high}%
  \BibitemOpen
  \bibfield  {author} {\bibinfo {author} {\bibfnamefont {Claudia}\ \bibnamefont {Piccinini}}, \bibinfo {author} {\bibfnamefont {Athanasios}\ \bibnamefont {Paralikis}}, \bibinfo {author} {\bibfnamefont {Jos{\'e}~Ferreira}\ \bibnamefont {Neto}}, \bibinfo {author} {\bibfnamefont {Abdulmalik~A}\ \bibnamefont {Madigawa}}, \bibinfo {author} {\bibfnamefont {Pawe{\l}}\ \bibnamefont {Wyborski}}, \bibinfo {author} {\bibfnamefont {Vikas}\ \bibnamefont {Remesh}}, \bibinfo {author} {\bibfnamefont {Luca}\ \bibnamefont {Vannucci}}, \bibinfo {author} {\bibfnamefont {Niels}\ \bibnamefont {Gregersen}}, \ and\ \bibinfo {author} {\bibfnamefont {Battulga}\ \bibnamefont {Munkhbat}},\ }\bibfield  {title} {\enquote {\bibinfo {title} {High-purity and stable single-photon emission in bilayer {WSe\(_2\)} via phonon-assisted excitation},}\ }\href@noop {} {\bibfield  {journal} {\bibinfo  {journal} {Communications Physics}\ }\textbf {\bibinfo {volume} {8}},\ \bibinfo {pages} {158} (\bibinfo {year} {2025})}\BibitemShut {NoStop}%
\bibitem [{\citenamefont {Tonndorf}\ \emph {et~al.}(2015)\citenamefont {Tonndorf}, \citenamefont {Schmidt}, \citenamefont {Schneider}, \citenamefont {Kern}, \citenamefont {Buscema}, \citenamefont {Steele}, \citenamefont {Castellanos-Gomez}, \citenamefont {van~der Zant}, \citenamefont {Michaelis~de Vasconcellos},\ and\ \citenamefont {Bratschitsch}}]{tonndorf2015single}%
  \BibitemOpen
  \bibfield  {author} {\bibinfo {author} {\bibfnamefont {Philipp}\ \bibnamefont {Tonndorf}}, \bibinfo {author} {\bibfnamefont {Robert}\ \bibnamefont {Schmidt}}, \bibinfo {author} {\bibfnamefont {Robert}\ \bibnamefont {Schneider}}, \bibinfo {author} {\bibfnamefont {Johannes}\ \bibnamefont {Kern}}, \bibinfo {author} {\bibfnamefont {Michele}\ \bibnamefont {Buscema}}, \bibinfo {author} {\bibfnamefont {Gary~A.}\ \bibnamefont {Steele}}, \bibinfo {author} {\bibfnamefont {Andres}\ \bibnamefont {Castellanos-Gomez}}, \bibinfo {author} {\bibfnamefont {Herre S.~J.}\ \bibnamefont {van~der Zant}}, \bibinfo {author} {\bibfnamefont {Steffen}\ \bibnamefont {Michaelis~de Vasconcellos}}, \ and\ \bibinfo {author} {\bibfnamefont {Rudolf}\ \bibnamefont {Bratschitsch}},\ }\bibfield  {title} {\enquote {\bibinfo {title} {Single-photon emission from localized excitons in an atomically thin semiconductor},}\ }\href@noop {} {\bibfield  {journal} {\bibinfo  {journal} {Optica}\ }\textbf {\bibinfo {volume} {2}},\ \bibinfo {pages}
  {347--352} (\bibinfo {year} {2015})}\BibitemShut {NoStop}%
\bibitem [{\citenamefont {Branny}\ \emph {et~al.}(2017)\citenamefont {Branny}, \citenamefont {Kumar}, \citenamefont {Proux},\ and\ \citenamefont {Gerardot}}]{branny2017deterministic}%
  \BibitemOpen
  \bibfield  {author} {\bibinfo {author} {\bibfnamefont {Artur}\ \bibnamefont {Branny}}, \bibinfo {author} {\bibfnamefont {Santosh}\ \bibnamefont {Kumar}}, \bibinfo {author} {\bibfnamefont {Rapha{\"e}l}\ \bibnamefont {Proux}}, \ and\ \bibinfo {author} {\bibfnamefont {Brian~D.}\ \bibnamefont {Gerardot}},\ }\bibfield  {title} {\enquote {\bibinfo {title} {Deterministic strain-induced arrays of quantum emitters in a two-dimensional semiconductor},}\ }\href@noop {} {\bibfield  {journal} {\bibinfo  {journal} {Nature Communications}\ }\textbf {\bibinfo {volume} {8}},\ \bibinfo {pages} {15053} (\bibinfo {year} {2017})}\BibitemShut {NoStop}%
\bibitem [{\citenamefont {Kim}\ \emph {et~al.}(2019)\citenamefont {Kim}, \citenamefont {Moon}, \citenamefont {Noh}, \citenamefont {Lee},\ and\ \citenamefont {Kim}}]{kim2019position}%
  \BibitemOpen
  \bibfield  {author} {\bibinfo {author} {\bibfnamefont {Hyoju}\ \bibnamefont {Kim}}, \bibinfo {author} {\bibfnamefont {Jong~Sung}\ \bibnamefont {Moon}}, \bibinfo {author} {\bibfnamefont {Gichang}\ \bibnamefont {Noh}}, \bibinfo {author} {\bibfnamefont {Jieun}\ \bibnamefont {Lee}}, \ and\ \bibinfo {author} {\bibfnamefont {Je-Hyung}\ \bibnamefont {Kim}},\ }\bibfield  {title} {\enquote {\bibinfo {title} {Position and frequency control of strain-induced quantum emitters in {WSe\(_2\)} monolayers},}\ }\href@noop {} {\bibfield  {journal} {\bibinfo  {journal} {Nano Letters}\ }\textbf {\bibinfo {volume} {19}},\ \bibinfo {pages} {7534--7539} (\bibinfo {year} {2019})}\BibitemShut {NoStop}%
\bibitem [{\citenamefont {Daveau}\ \emph {et~al.}(2020)\citenamefont {Daveau}, \citenamefont {Vandekerckhove}, \citenamefont {Mukherjee}, \citenamefont {Wang}, \citenamefont {Shan}, \citenamefont {Mak}, \citenamefont {Vamivakas},\ and\ \citenamefont {Fuchs}}]{daveau2020spectral}%
  \BibitemOpen
  \bibfield  {author} {\bibinfo {author} {\bibfnamefont {Rapha{\"e}l~S.}\ \bibnamefont {Daveau}}, \bibinfo {author} {\bibfnamefont {Tom}\ \bibnamefont {Vandekerckhove}}, \bibinfo {author} {\bibfnamefont {Arunabh}\ \bibnamefont {Mukherjee}}, \bibinfo {author} {\bibfnamefont {Zefang}\ \bibnamefont {Wang}}, \bibinfo {author} {\bibfnamefont {Jie}\ \bibnamefont {Shan}}, \bibinfo {author} {\bibfnamefont {Kin~Fai}\ \bibnamefont {Mak}}, \bibinfo {author} {\bibfnamefont {A.~Nick}\ \bibnamefont {Vamivakas}}, \ and\ \bibinfo {author} {\bibfnamefont {Gregory~D.}\ \bibnamefont {Fuchs}},\ }\bibfield  {title} {\enquote {\bibinfo {title} {Spectral and spatial isolation of single tungsten diselenide quantum emitters using hexagonal boron nitride wrinkles},}\ }\href@noop {} {\bibfield  {journal} {\bibinfo  {journal} {APL Photonics}\ }\textbf {\bibinfo {volume} {5}} (\bibinfo {year} {2020})}\BibitemShut {NoStop}%
\bibitem [{\citenamefont {Androulidakis}\ \emph {et~al.}(2018)\citenamefont {Androulidakis}, \citenamefont {Zhang}, \citenamefont {Robertson},\ and\ \citenamefont {Tawfick}}]{androulidakis2018tailoring}%
  \BibitemOpen
  \bibfield  {author} {\bibinfo {author} {\bibfnamefont {Charalampos}\ \bibnamefont {Androulidakis}}, \bibinfo {author} {\bibfnamefont {Kaihao}\ \bibnamefont {Zhang}}, \bibinfo {author} {\bibfnamefont {Matthew}\ \bibnamefont {Robertson}}, \ and\ \bibinfo {author} {\bibfnamefont {Sameh}\ \bibnamefont {Tawfick}},\ }\bibfield  {title} {\enquote {\bibinfo {title} {Tailoring the mechanical properties of 2d materials and heterostructures},}\ }\href@noop {} {\bibfield  {journal} {\bibinfo  {journal} {2D Materials}\ }\textbf {\bibinfo {volume} {5}},\ \bibinfo {pages} {032005} (\bibinfo {year} {2018})}\BibitemShut {NoStop}%
\bibitem [{\citenamefont {So}\ \emph {et~al.}(2021)\citenamefont {So}, \citenamefont {Jeong}, \citenamefont {Lee}, \citenamefont {Kim}, \citenamefont {Lee}, \citenamefont {Huh}, \citenamefont {Kim}, \citenamefont {Choi}, \citenamefont {Kim}, \citenamefont {Kim} \emph {et~al.}}]{so2021polarization}%
  \BibitemOpen
  \bibfield  {author} {\bibinfo {author} {\bibfnamefont {Jae-Pil}\ \bibnamefont {So}}, \bibinfo {author} {\bibfnamefont {Kwang-Yong}\ \bibnamefont {Jeong}}, \bibinfo {author} {\bibfnamefont {Jung~Min}\ \bibnamefont {Lee}}, \bibinfo {author} {\bibfnamefont {Kyoung-Ho}\ \bibnamefont {Kim}}, \bibinfo {author} {\bibfnamefont {Soon-Jae}\ \bibnamefont {Lee}}, \bibinfo {author} {\bibfnamefont {Woong}\ \bibnamefont {Huh}}, \bibinfo {author} {\bibfnamefont {Ha-Reem}\ \bibnamefont {Kim}}, \bibinfo {author} {\bibfnamefont {Jae-Hyuck}\ \bibnamefont {Choi}}, \bibinfo {author} {\bibfnamefont {Jin~Myung}\ \bibnamefont {Kim}}, \bibinfo {author} {\bibfnamefont {Yoon~Seok}\ \bibnamefont {Kim}},  \emph {et~al.},\ }\bibfield  {title} {\enquote {\bibinfo {title} {Polarization control of deterministic single-photon emitters in monolayer {WSe\(_2\)}},}\ }\href@noop {} {\bibfield  {journal} {\bibinfo  {journal} {Nano Letters}\ }\textbf {\bibinfo {volume} {21}},\ \bibinfo {pages} {1546--1554} (\bibinfo {year} {2021})}\BibitemShut
  {NoStop}%
\bibitem [{\citenamefont {Iff}\ \emph {et~al.}(2021)\citenamefont {Iff}, \citenamefont {Buchinger}, \citenamefont {Mocza{\l}a-Dusanowska}, \citenamefont {Kamp}, \citenamefont {Betzold}, \citenamefont {Davanco}, \citenamefont {Srinivasan}, \citenamefont {Tongay}, \citenamefont {Ant{\'o}n-Solanas}, \citenamefont {H{\"o}fling} \emph {et~al.}}]{iff2021purcell}%
  \BibitemOpen
  \bibfield  {author} {\bibinfo {author} {\bibfnamefont {Oliver}\ \bibnamefont {Iff}}, \bibinfo {author} {\bibfnamefont {Quirin}\ \bibnamefont {Buchinger}}, \bibinfo {author} {\bibfnamefont {Magdalena}\ \bibnamefont {Mocza{\l}a-Dusanowska}}, \bibinfo {author} {\bibfnamefont {Martin}\ \bibnamefont {Kamp}}, \bibinfo {author} {\bibfnamefont {Simon}\ \bibnamefont {Betzold}}, \bibinfo {author} {\bibfnamefont {Marcelo}\ \bibnamefont {Davanco}}, \bibinfo {author} {\bibfnamefont {Kartik}\ \bibnamefont {Srinivasan}}, \bibinfo {author} {\bibfnamefont {Sefaattin}\ \bibnamefont {Tongay}}, \bibinfo {author} {\bibfnamefont {Carlos}\ \bibnamefont {Ant{\'o}n-Solanas}}, \bibinfo {author} {\bibfnamefont {Sven}\ \bibnamefont {H{\"o}fling}},  \emph {et~al.},\ }\bibfield  {title} {\enquote {\bibinfo {title} {Purcell-enhanced single photon source based on a deterministically placed {WSe\(_2\)} monolayer quantum dot in a circular bragg grating cavity},}\ }\href@noop {} {\bibfield  {journal} {\bibinfo  {journal} {Nano Letters}\
  }\textbf {\bibinfo {volume} {21}},\ \bibinfo {pages} {4715--4720} (\bibinfo {year} {2021})}\BibitemShut {NoStop}%
\bibitem [{\citenamefont {Jahn}\ \emph {et~al.}(2015)\citenamefont {Jahn}, \citenamefont {Munsch}, \citenamefont {B{\'e}guin}, \citenamefont {Kuhlmann}, \citenamefont {Renggli}, \citenamefont {Huo}, \citenamefont {Ding}, \citenamefont {Trotta}, \citenamefont {Reindl}, \citenamefont {Schmidt} \emph {et~al.}}]{jahn2015artificial}%
  \BibitemOpen
  \bibfield  {author} {\bibinfo {author} {\bibfnamefont {Jan-Philipp}\ \bibnamefont {Jahn}}, \bibinfo {author} {\bibfnamefont {Mathieu}\ \bibnamefont {Munsch}}, \bibinfo {author} {\bibfnamefont {Lucas}\ \bibnamefont {B{\'e}guin}}, \bibinfo {author} {\bibfnamefont {Andreas~V.}\ \bibnamefont {Kuhlmann}}, \bibinfo {author} {\bibfnamefont {Martina}\ \bibnamefont {Renggli}}, \bibinfo {author} {\bibfnamefont {Yongheng}\ \bibnamefont {Huo}}, \bibinfo {author} {\bibfnamefont {Fei}\ \bibnamefont {Ding}}, \bibinfo {author} {\bibfnamefont {Rinaldo}\ \bibnamefont {Trotta}}, \bibinfo {author} {\bibfnamefont {Marcus}\ \bibnamefont {Reindl}}, \bibinfo {author} {\bibfnamefont {Oliver~G.}\ \bibnamefont {Schmidt}},  \emph {et~al.},\ }\bibfield  {title} {\enquote {\bibinfo {title} {An artificial rb atom in a semiconductor with lifetime-limited linewidth},}\ }\href@noop {} {\bibfield  {journal} {\bibinfo  {journal} {Physical Review B}\ }\textbf {\bibinfo {volume} {92}},\ \bibinfo {pages} {245439} (\bibinfo {year}
  {2015})}\BibitemShut {NoStop}%
\bibitem [{\citenamefont {Zhai}\ \emph {et~al.}(2020)\citenamefont {Zhai}, \citenamefont {L{\"o}bl}, \citenamefont {Nguyen}, \citenamefont {Ritzmann}, \citenamefont {Javadi}, \citenamefont {Spinnler}, \citenamefont {Wieck}, \citenamefont {Ludwig},\ and\ \citenamefont {Warburton}}]{zhai2020low}%
  \BibitemOpen
  \bibfield  {author} {\bibinfo {author} {\bibfnamefont {Liang}\ \bibnamefont {Zhai}}, \bibinfo {author} {\bibfnamefont {Matthias~C.}\ \bibnamefont {L{\"o}bl}}, \bibinfo {author} {\bibfnamefont {Giang~N.}\ \bibnamefont {Nguyen}}, \bibinfo {author} {\bibfnamefont {Julian}\ \bibnamefont {Ritzmann}}, \bibinfo {author} {\bibfnamefont {Alisa}\ \bibnamefont {Javadi}}, \bibinfo {author} {\bibfnamefont {Clemens}\ \bibnamefont {Spinnler}}, \bibinfo {author} {\bibfnamefont {Andreas~D.}\ \bibnamefont {Wieck}}, \bibinfo {author} {\bibfnamefont {Arne}\ \bibnamefont {Ludwig}}, \ and\ \bibinfo {author} {\bibfnamefont {Richard~J.}\ \bibnamefont {Warburton}},\ }\bibfield  {title} {\enquote {\bibinfo {title} {Low-noise gaas quantum dots for quantum photonics},}\ }\href@noop {} {\bibfield  {journal} {\bibinfo  {journal} {Nature Communications}\ }\textbf {\bibinfo {volume} {11}},\ \bibinfo {pages} {4745} (\bibinfo {year} {2020})}\BibitemShut {NoStop}%
\bibitem [{\citenamefont {Koperski}\ \emph {et~al.}(2015)\citenamefont {Koperski}, \citenamefont {Nogajewski}, \citenamefont {Arora}, \citenamefont {Cherkez}, \citenamefont {Mallet}, \citenamefont {Veuillen}, \citenamefont {Marcus}, \citenamefont {Kossacki},\ and\ \citenamefont {Potemski}}]{koperski2015single}%
  \BibitemOpen
  \bibfield  {author} {\bibinfo {author} {\bibfnamefont {Maciej}\ \bibnamefont {Koperski}}, \bibinfo {author} {\bibfnamefont {K.}~\bibnamefont {Nogajewski}}, \bibinfo {author} {\bibfnamefont {Ashish}\ \bibnamefont {Arora}}, \bibinfo {author} {\bibfnamefont {V.}~\bibnamefont {Cherkez}}, \bibinfo {author} {\bibfnamefont {Paul}\ \bibnamefont {Mallet}}, \bibinfo {author} {\bibfnamefont {J.-Y.}\ \bibnamefont {Veuillen}}, \bibinfo {author} {\bibfnamefont {J.}~\bibnamefont {Marcus}}, \bibinfo {author} {\bibfnamefont {Piotr}\ \bibnamefont {Kossacki}}, \ and\ \bibinfo {author} {\bibfnamefont {M}~\bibnamefont {Potemski}},\ }\bibfield  {title} {\enquote {\bibinfo {title} {Single photon emitters in exfoliated {WSe\(_2\)} structures},}\ }\href@noop {} {\bibfield  {journal} {\bibinfo  {journal} {Nature Nanotechnology}\ }\textbf {\bibinfo {volume} {10}},\ \bibinfo {pages} {503--506} (\bibinfo {year} {2015})}\BibitemShut {NoStop}%
\bibitem [{\citenamefont {Al{\'e}n}\ \emph {et~al.}(2007)\citenamefont {Al{\'e}n}, \citenamefont {Bosch}, \citenamefont {Granados}, \citenamefont {Mart{\'\i}nez-Pastor}, \citenamefont {Garc{\'\i}a},\ and\ \citenamefont {Gonz{\'a}lez}}]{alen2007oscillator}%
  \BibitemOpen
  \bibfield  {author} {\bibinfo {author} {\bibfnamefont {Benito}\ \bibnamefont {Al{\'e}n}}, \bibinfo {author} {\bibfnamefont {Jos{\'e}}\ \bibnamefont {Bosch}}, \bibinfo {author} {\bibfnamefont {Daniel}\ \bibnamefont {Granados}}, \bibinfo {author} {\bibfnamefont {Juan}\ \bibnamefont {Mart{\'\i}nez-Pastor}}, \bibinfo {author} {\bibfnamefont {Jorge~M}\ \bibnamefont {Garc{\'\i}a}}, \ and\ \bibinfo {author} {\bibfnamefont {Luisa}\ \bibnamefont {Gonz{\'a}lez}},\ }\bibfield  {title} {\enquote {\bibinfo {title} {Oscillator strength reduction induced by external electric fields in self-assembled quantum dots and rings},}\ }\href@noop {} {\bibfield  {journal} {\bibinfo  {journal} {Physical Review B}\ }\textbf {\bibinfo {volume} {75}},\ \bibinfo {pages} {045319} (\bibinfo {year} {2007})}\BibitemShut {NoStop}%
\bibitem [{\citenamefont {Cadiz}\ \emph {et~al.}(2017)\citenamefont {Cadiz}, \citenamefont {Courtade}, \citenamefont {Robert}, \citenamefont {Wang}, \citenamefont {Shen}, \citenamefont {Cai}, \citenamefont {Taniguchi}, \citenamefont {Watanabe}, \citenamefont {Carrere}, \citenamefont {Lagarde} \emph {et~al.}}]{cadiz2017excitonic}%
  \BibitemOpen
  \bibfield  {author} {\bibinfo {author} {\bibfnamefont {Fabian}\ \bibnamefont {Cadiz}}, \bibinfo {author} {\bibfnamefont {Emmanuel}\ \bibnamefont {Courtade}}, \bibinfo {author} {\bibfnamefont {C{\'e}dric}\ \bibnamefont {Robert}}, \bibinfo {author} {\bibfnamefont {Gang}\ \bibnamefont {Wang}}, \bibinfo {author} {\bibfnamefont {Yuxia}\ \bibnamefont {Shen}}, \bibinfo {author} {\bibfnamefont {Hui}\ \bibnamefont {Cai}}, \bibinfo {author} {\bibfnamefont {Takashi}\ \bibnamefont {Taniguchi}}, \bibinfo {author} {\bibfnamefont {Kenji}\ \bibnamefont {Watanabe}}, \bibinfo {author} {\bibfnamefont {Helene}\ \bibnamefont {Carrere}}, \bibinfo {author} {\bibfnamefont {Delphine}\ \bibnamefont {Lagarde}},  \emph {et~al.},\ }\bibfield  {title} {\enquote {\bibinfo {title} {Excitonic linewidth approaching the homogeneous limit in {MoS\(_2\)}-based van der waals heterostructures},}\ }\href@noop {} {\bibfield  {journal} {\bibinfo  {journal} {Physical Review X}\ }\textbf {\bibinfo {volume} {7}},\ \bibinfo {pages} {021026} (\bibinfo
  {year} {2017})}\BibitemShut {NoStop}%
\bibitem [{\citenamefont {Chow}\ \emph {et~al.}(2017)\citenamefont {Chow}, \citenamefont {Yu}, \citenamefont {Jones}, \citenamefont {Yan}, \citenamefont {Mandrus}, \citenamefont {Taniguchi}, \citenamefont {Watanabe}, \citenamefont {Yao},\ and\ \citenamefont {Xu}}]{chow2017unusual}%
  \BibitemOpen
  \bibfield  {author} {\bibinfo {author} {\bibfnamefont {Colin~M.}\ \bibnamefont {Chow}}, \bibinfo {author} {\bibfnamefont {Hongyi}\ \bibnamefont {Yu}}, \bibinfo {author} {\bibfnamefont {Aaron~M.}\ \bibnamefont {Jones}}, \bibinfo {author} {\bibfnamefont {Jiaqiang}\ \bibnamefont {Yan}}, \bibinfo {author} {\bibfnamefont {David~G.}\ \bibnamefont {Mandrus}}, \bibinfo {author} {\bibfnamefont {Takashi}\ \bibnamefont {Taniguchi}}, \bibinfo {author} {\bibfnamefont {Kenji}\ \bibnamefont {Watanabe}}, \bibinfo {author} {\bibfnamefont {Wang}\ \bibnamefont {Yao}}, \ and\ \bibinfo {author} {\bibfnamefont {Xiaodong}\ \bibnamefont {Xu}},\ }\bibfield  {title} {\enquote {\bibinfo {title} {Unusual exciton--phonon interactions at van der waals engineered interfaces},}\ }\href@noop {} {\bibfield  {journal} {\bibinfo  {journal} {Nano Letters}\ }\textbf {\bibinfo {volume} {17}},\ \bibinfo {pages} {1194--1199} (\bibinfo {year} {2017})}\BibitemShut {NoStop}%
\bibitem [{\citenamefont {Saleh}\ and\ \citenamefont {Teich}(2019)}]{saleh2019fundamentals}%
  \BibitemOpen
  \bibfield  {author} {\bibinfo {author} {\bibfnamefont {Bahaa E.~A.}\ \bibnamefont {Saleh}}\ and\ \bibinfo {author} {\bibfnamefont {Malvin~Carl}\ \bibnamefont {Teich}},\ }\href@noop {} {\emph {\bibinfo {title} {Fundamentals of photonics, 2 volume set}}}\ (\bibinfo  {publisher} {John Wiley \& Sons},\ \bibinfo {year} {2019})\BibitemShut {NoStop}%
\bibitem [{\citenamefont {Karli}\ \emph {et~al.}(2022)\citenamefont {Karli}, \citenamefont {Kappe}, \citenamefont {Remesh}, \citenamefont {Bracht}, \citenamefont {M{\"u}nzberg}, \citenamefont {Covre~da Silva}, \citenamefont {Seidelmann}, \citenamefont {Axt}, \citenamefont {Rastelli}, \citenamefont {Reiter} \emph {et~al.}}]{karli2022super}%
  \BibitemOpen
  \bibfield  {author} {\bibinfo {author} {\bibfnamefont {Yusuf}\ \bibnamefont {Karli}}, \bibinfo {author} {\bibfnamefont {Florian}\ \bibnamefont {Kappe}}, \bibinfo {author} {\bibfnamefont {Vikas}\ \bibnamefont {Remesh}}, \bibinfo {author} {\bibfnamefont {Thomas~K}\ \bibnamefont {Bracht}}, \bibinfo {author} {\bibfnamefont {Julian}\ \bibnamefont {M{\"u}nzberg}}, \bibinfo {author} {\bibfnamefont {Saimon}\ \bibnamefont {Covre~da Silva}}, \bibinfo {author} {\bibfnamefont {Tim}\ \bibnamefont {Seidelmann}}, \bibinfo {author} {\bibfnamefont {Vollrath~Martin}\ \bibnamefont {Axt}}, \bibinfo {author} {\bibfnamefont {Armando}\ \bibnamefont {Rastelli}}, \bibinfo {author} {\bibfnamefont {Doris~E.}\ \bibnamefont {Reiter}},  \emph {et~al.},\ }\bibfield  {title} {\enquote {\bibinfo {title} {Super scheme in action: Experimental demonstration of red-detuned excitation of a quantum emitter},}\ }\href@noop {} {\bibfield  {journal} {\bibinfo  {journal} {Nano Letters}\ }\textbf {\bibinfo {volume} {22}},\ \bibinfo {pages}
  {6567--6572} (\bibinfo {year} {2022})}\BibitemShut {NoStop}%
\bibitem [{\citenamefont {Kumar}\ \emph {et~al.}(2016)\citenamefont {Kumar}, \citenamefont {Brot{\'o}ns-Gisbert}, \citenamefont {Al-Khuzheyri}, \citenamefont {Branny}, \citenamefont {Ballesteros-Garcia}, \citenamefont {S{\'a}nchez-Royo},\ and\ \citenamefont {Gerardot}}]{kumar2016resonant}%
  \BibitemOpen
  \bibfield  {author} {\bibinfo {author} {\bibfnamefont {Santosh}\ \bibnamefont {Kumar}}, \bibinfo {author} {\bibfnamefont {Mauro}\ \bibnamefont {Brot{\'o}ns-Gisbert}}, \bibinfo {author} {\bibfnamefont {Rima}\ \bibnamefont {Al-Khuzheyri}}, \bibinfo {author} {\bibfnamefont {Artur}\ \bibnamefont {Branny}}, \bibinfo {author} {\bibfnamefont {Guillem}\ \bibnamefont {Ballesteros-Garcia}}, \bibinfo {author} {\bibfnamefont {Juan~F.}\ \bibnamefont {S{\'a}nchez-Royo}}, \ and\ \bibinfo {author} {\bibfnamefont {Brian~D.}\ \bibnamefont {Gerardot}},\ }\bibfield  {title} {\enquote {\bibinfo {title} {Resonant laser spectroscopy of localized excitons in monolayer {WSe\(_2\)}},}\ }\href@noop {} {\bibfield  {journal} {\bibinfo  {journal} {Optica}\ }\textbf {\bibinfo {volume} {3}},\ \bibinfo {pages} {882--886} (\bibinfo {year} {2016})}\BibitemShut {NoStop}%
\bibitem [{\citenamefont {Castellanos-Gomez}\ \emph {et~al.}(2014)\citenamefont {Castellanos-Gomez}, \citenamefont {Buscema}, \citenamefont {Molenaar}, \citenamefont {Singh}, \citenamefont {Janssen}, \citenamefont {Van Der~Zant},\ and\ \citenamefont {Steele}}]{castellanos2014deterministic}%
  \BibitemOpen
  \bibfield  {author} {\bibinfo {author} {\bibfnamefont {Andres}\ \bibnamefont {Castellanos-Gomez}}, \bibinfo {author} {\bibfnamefont {Michele}\ \bibnamefont {Buscema}}, \bibinfo {author} {\bibfnamefont {Rianda}\ \bibnamefont {Molenaar}}, \bibinfo {author} {\bibfnamefont {Vibhor}\ \bibnamefont {Singh}}, \bibinfo {author} {\bibfnamefont {Laurens}\ \bibnamefont {Janssen}}, \bibinfo {author} {\bibfnamefont {Herre S.~J.}\ \bibnamefont {Van Der~Zant}}, \ and\ \bibinfo {author} {\bibfnamefont {Gary~A.}\ \bibnamefont {Steele}},\ }\bibfield  {title} {\enquote {\bibinfo {title} {Deterministic transfer of two-dimensional materials by all-dry viscoelastic stamping},}\ }\href@noop {} {\bibfield  {journal} {\bibinfo  {journal} {2D Materials}\ }\textbf {\bibinfo {volume} {1}},\ \bibinfo {pages} {011002} (\bibinfo {year} {2014})}\BibitemShut {NoStop}%
\end{thebibliography}%
